\documentclass[12pt]{article}

\usepackage{graphicx}
\usepackage{amssymb}
\usepackage{bm}
\usepackage{epstopdf}
\title{Theory of Graphene Raman Scattering	}
\author
{Eric J. Heller, $^{1,2\ast}$ Yuan Yang,$^2$ Lucas Kocia,$^2$ Wei Chen,$^1$ \\Shiang Fang,$^1$ Mario Borunda,$^3$ Efthimios Kaxiras,$^1$\\
\\
\normalsize{$^{1}$Department of Physics,  Harvard University, Cambridge,  MA  02138, USA}\\
\normalsize{$^{2}$Department of Chemistry and Chemical Biology, Harvard University, Cambridge,  MA  02138, USA}\\
\normalsize{$^{1}$Department of Physics, Oklahoma State University, Stillwater, OK 74078, USA}\\
\\
\normalsize{$^\ast$To whom correspondence should be addressed; E-mail:  heller@physics.harvard.edu.}
}

 \date{}
\begin{document}
 \maketitle

  \begin{abstract}
 Raman scattering plays a key  role in unraveling the quantum dynamics of graphene, perhaps the most promising material of recent times.  It is crucial to correctly interpret the meaning of the spectra.  It is therefore very surprising that the widely accepted understanding of Raman scattering, i.e.  Kramers-Heisenberg-Dirac theory,  has never been applied to graphene. Doing so here, a remarkable  mechanism we term``transition sliding'' is uncovered, explaining the   uncommon brightness of overtones in graphene. 
Graphene's  dispersive and fixed Raman bands,  missing bands,   defect density and laser frequency dependence  of band intensities, widths of overtone bands,   Stokes, anti-Stokes anomalies,   and other  known properties emerge  simply and directly.   \end{abstract}

 \section {Introduction}

  The unique properties of graphene and related systems have propelled it to a high level of   interest for more than a decade. Raman scattering is perhaps the key  window on graphene's quantum properties, yet the   crucial  aspects of the spectrum of graphene, carbon nanotubes,  and graphite Raman spectra  have been a subject of much controversy, in some cases for decades. 
                                                                                                                                                                                                                                                                                                                                                                                                                                        
The delay in applying    Kramers-Heisenberg-Dirac theory (KHD)  to graphene is attributable to the rise of a non-traditional, 4th order perturbative   Raman model (KHD is 2nd order) called ``double resonance'' (DR), which  appeared 15 or so years ago\cite{reich},  rapidly gaining   wide                                                                                                                                                                                                                                                       acceptance  exclusively in the conjugated carbon Raman community.  Far from being a variant of KHD tuned to graphene, DR is starkly incompatible with KHD.    DR exclusively relies on  post-photoabsorption  inelastic electronic scattering  to create phonons, requiring introduction of two additional orders of electron-phonon perturbation theory, making DR overall fourth order.   These make no appearance in KHD, which  cleanly produces phonons free of Pauli blocking by two other mechanisms, overall in second order perturbation theory.  

Search engines quickly reveal that none of the 3000+ published contributions  developing DR or using it to interpret experiments has     mentioned  the much older and more established KHD Raman formalism, in the last 15 years.   No   rationale  was ever given as to why a radically different, non KHD formalism was needed. On the face of it, even without regard to the theory given in this paper, these facts leave the DR approach  seriously unexamined and exposed.  We show here that unfortunately DR was in fact a wrong turn on the way to understanding  graphene quantum dynamics and Raman spectroscopy.

  When  KHD is applied to graphene, phonons are produced exclusively and instantly at the moment of photoabsorption or photoemission by the long neglected (in condensed matter physics) nuclear coordinate dependence of the transition moment.   Graphene's Raman spectroscopy  falls quickly and naturally into place, providing new insights and predictions along the way.



 The present graphene study has a strong precedent in another conjugated conducting polymer. In a recent  study \cite{heller2015raman},   the long-standing mysteries of the  dispersive polyacetylene Raman spectrum yielded to   KHD theory, extended to include 1-D crystal structure, defects, and electron and phonon dispersion relations.  All these arise from  within Born-Oppenheimer theory and KHD Raman scattering theory.  The coordinate dependence of the transition moment plays a key role, since geometry does not change when the electron-hole pairs are created in the extended conjugated system.

 These principles are the starting point for surprising new insights that  graphene has in store. For example, ``transition sliding'',  not possible in a DR picture,   is responsible for  the brightness of overtone bands, and explains their impurity and doping dependence. Transition sliding does not apply to polyacetylene because there are no Dirac cones; indeed polyacetylene  overtones are  correspondingly weak.

       \vskip .07in
\section{Virtual processes in KHD \textit  {versus} DR}
  \vskip .07in
    
 DR makes heavy use of virtual states that are foreign to ABO.   ABO is certainly not exact (see next section), so for example its predicted  electronic band structure  is not exact. Still it is surprising to see virtual states  freely used in DR that do not exist   on the  electronic bands, or virtual states that violate momentum conservation, or   even   violate the  Pauli exclusion principle\cite{ferrariNat}.      
   
  If one accepts ABO as at least close to the truth,  no such virtual states can  arise.  All the terms in the KHD sum are momentum conserving, Pauli principle obeying states lying exactly on Dirac cones or other parts of the electronic dispersion surfaces. KHD is based  on second order time dependent light-matter perturbation theory applied to Born-Oppenheimer states. Only the energy, conjugate to time, can become virtual, as familiar in off-resonant Raman scattering.     Even ``temporary'' violation of momentum conservation   is   forbidden in KHD, since momentum violating terms in the sum in equation  \ref{bopp} give vanishing matrix elements, whether they are off-resonance or not.

   \vskip .07in
\section{Does the Born-Oppenheimer Approximation ``Break Down'' for Graphene?}
  \vskip .07in
  
  KHD relies heavily on the Born-Oppenheimer approximation, as does much of condensed matter physics.   
   Recently, the  ``headlines'' in a highly cited  paper left the  impression   that  the ABO fails  particularly in graphene. 
 The article was excellent, article with the  provocative title ``Breakdown of 
  the adiabatic Born-Oppenheimer approximation in graphene'' \cite{breakdown}, published in Nature Materials  with over 800 citations,  we find the statements ``... ABO has proved effective for the accurate determination of chemical reactions, molecular dynamics and phonon frequencies  in a wide range of metallic systems. Here, we show that ABO fails in graphene.'' and later ``Quite remarkably, the ABO fails in graphene''\cite{breakdown}, see also \cite{Ferrari_ePh}. Such  statements read like an announcement of  catastrophic  failure of ABO in graphene.    Several times in the course of explaining the present work, one of the authors has been asked, ``but didn't you know that the     Born-Oppenheimer approximation fails in graphene?"

Of course the Born-Oppenheimer approximation fail in graphene, and in every other system it has been applied to in the last 90 years, in specific regimes and situations.  It is, after all, an approximation. Reference~\cite{breakdown}   claims nothing so drastic as a catastrophic breakdown once one reads deeper. Instead it makes   important points about a  stiffening of the G mode with electron density near the Dirac point involving  the Kohn anomaly,  missed by ABO.  By their nature Kohn anomalies involve rapid changes of electronic structure with small nuclear configuration changes, a warning sign of potential ABO breakdown.  In this paper we do not attempt to correct for the Kohn anomalies and their effect on phonon modes and mode frequencies.  Electron-hole pairs created in Raman studies are normally well away from the Dirac point and orbitals affected by  Kohn anomalies.
  
  The   Born-Oppenheimer approximation,   proposed in 1927,  is the  basis of crystal structure, electronic band structure, and phonons and their band structure.  It is quite arguably \textit{the} main pillar  of condensed matter physics.  Surely it needs monitoring and corrections, but without it we have a structureless pea soup of electrons and nuclei, rather than crystals and solids.   
  
  The  ABO succeeds in the large, that is the point.     Recall that within ABO, if the nuclei return to a prior configuration, the electrons do also.  This is  strictly incompatible with inelastic electron-phonon scattering having taken place in the meantime.  Inelastic electron-phonon scattering is nonetheless a valid non-adiabatic correction to ABO.  However, electron-phonon scattering would appear to Pauli block the affected electrons, immediately  rendering them silent in Raman scattering. 
  
We have been given a late opportunity to use KHD theory the first time in graphene. ABO is vastly more established  than DR, and astonishingly accurate most of the time.  It much simpler to execute, and lower order in perturbation theory. We hope the community is curious as to how it performs.
  
  .  

     \vskip .07in
\section{Kramers-Heisenberg-Dirac Theory}
  \vskip .07in
  \label{KHDT}
 We provide a   review of KHD here; facilitating the developments in succeeding sections.
 Just before the dawn of quantum mechanics, in 1925  Kramers and Heisenberg published a correspondence principle account of Raman scattering \cite{kramers1925streuung,soo}, which Dirac translated into quantum form in 1927 \cite{1927Dirac}.   The Kramers-Heisenberg-Dirac Theory of Raman scattering has been used  ever since to explain  more than half a million  Raman spectra  in a very wide range of systems, including very large conjugated hydrocarbons.  There is no reason  that removing hydrogens from carbon  materials should  cause KHD to catastrophically fail and require replacement by a theory based on different physics.

    The KHD formula  for the total Raman cross section $\Sigma$  reads, 
for incident frequency $\omega_I$ and polarization $\rho$,  scattered frequency  $\omega_s$ and polarization $\sigma$, between initial Born-Oppenheimer state $\vert i\rangle$ and final Born-Oppenheimer state $\vert f\rangle$ via intermediate 
 Born-Oppenheimer states $\vert n\rangle$ reads
 \begin{equation}
 \label{KHD}
 \Sigma_{i\to f}^{\rho,\sigma}  = \frac{8 \pi e^4 \omega_s^3\omega_I}{9 c^4} \left \vert \alpha_{i,f}^{\rho,\sigma}\right \vert^2;\ \  \alpha_{i,f}^{\rho,\sigma} =\frac{1}{\hbar}   \sum_{n}\left [  \frac{\langle f | {\bf D}^{\dagger,\sigma} | n \rangle \langle n | {\bf D}^\rho | i \rangle}{E_i - E_n + \hbar \omega_I - i \Gamma_n}+\frac{\langle f |{\bf D}^{\dagger,\sigma} | n \rangle \langle n | {\bf D}^\rho | i \rangle}{E_i + E_n + \hbar \omega_I + i  \Gamma_n}\ \right ],
\end{equation}
where $\Gamma_{n}$ is the damping factor for the excited state , accounting for events and degrees of freedom not explicitly represented. The transition moment operator ${\bf D}$ controls the first order perturbative matter-radiation coupling. Usually the second, non-resonant  term inside the square in equation~\ref{KHD} is  neglected, for simplicity. 

Making this more explicit, suppose with phonon coordinates  ${\boldsymbol \xi}$  and electron coordinates $\textbf{r}$,  we write
\begin{eqnarray}
\label{bopp}
\vert f\rangle &=& \vert\psi_{{\bf m}_f }^{B.O.}({\boldsymbol \xi},{\bf r}) \rangle = \vert\phi({\boldsymbol \xi};{\bf r})\rangle\vert \chi_{{\bf m}_f}({\boldsymbol \xi})\rangle,  \nonumber \\ \vert i\rangle  &=& \vert\psi_{{\bf m}_i }^{B.O.}({\boldsymbol \xi},{\bf r}) \rangle = \vert\phi({\boldsymbol \xi};{\bf r})\rangle\vert \chi_{{\bf m}_i}({\boldsymbol \xi})\rangle  \textrm{, and }   \nonumber \\  \vert n\rangle &=& \vert\psi_{{\bf q}_v{\bf q}_c,{\bf m}}^{B.O.}({\boldsymbol \xi},{\bf r}) \rangle =\vert\phi_{{\bf q}_v{\bf q}_c}({\boldsymbol \xi};{\bf r})\rangle\vert \chi_{{\bf m}_{vc}}({\boldsymbol \xi})\rangle.
\end{eqnarray}
$\vert\phi({\boldsymbol \xi};{\bf r})\rangle$ is the approximation to the Born-Oppenheimer $\pi$ electron ground state (we suppress the other electrons) based on a Slater determinant of  valence  electron spin orbitals; 
$\vert\phi_{{\bf q}_v{\bf q}_c}({\boldsymbol \xi};{\bf r})\rangle$ is an electron-hole pair  relative   to the ground state, with an electron in the conduction band orbital $\vert\phi_{ {\bf q}_c}({\boldsymbol \xi};{\bf r})\rangle$ and a hole in the valence band orbital $\vert\phi_{ {\bf q}_v}({\boldsymbol \xi};{\bf r})\rangle$. $\vert\phi_{{\bf q}_v{\bf q}_c}({\boldsymbol \xi};{\bf r})\rangle\vert \chi_{{\bf m}_{vc}}({\boldsymbol \xi})\rangle$ is a complete intermediate state, including the phonon wavefunction. $ {\bf q}_v $,  $ {\bf q}_c $ and ${\bf m_{vc}}$ may range quite freely, with the following remarks: (1) almost all   $ {\bf q}_v{\bf q}_c $, ${\bf m}_{vc}$ will  give vanishing matrix elements with the dipole operator, due to  momentum non-conservation (including the momentum of the phonons).(2) Some pairs $ {\bf q}_v{\bf q}_c $ with unmatched momentum $ {\bf q}_v \ne{\bf q}_c $ give non-vanishing matrix elements nonetheless because they are associated  with newly created   or destroyed phonons contained in $\vert \chi_{{\bf m}_{vc}}({\boldsymbol \xi})\rangle$   relative to the initial state,  conserving momentum.  Such states are the electron-hole-phonon triplets.  (3) Momentum is conserved   when elastic backscattering  is required to re-align the hole and particle (because of the kick given to the conduction electron by newly formed or destroyed phonons)	through the recoil of the whole sample, or laboratory, as is  familiar in 
  M\"ossbauer  scattering or elastic neutron scattering for example. This feature is embedded in equation~\ref{bopp} since the electronic orbitals   include the presence of impurities, edges, etc., which are part and parcel of ABO.  Even though the eigenfunction orbitals thus include the elastic backscattering as a boundary condition entirely within ABO, an electron  promoted with a given momentum becomes a linear superposition of such states, and still takes  time    to backscatter. In the energy domain, being off resonance in effect shortens the time and nearly removes the effects of the  backscattering.

 \vskip .07in
 \subsection{The transition moment operator} 
  \vskip .07in 
  
We may write
 \begin{equation}
\langle n | {\bf D}^\rho | i \rangle = \langle\chi_{{\bf m}_{vc}}({\boldsymbol \xi})\vert \langle\phi_{{\bf q}_v{\bf q}_c}({\boldsymbol \xi};{\bf r})\vert{\bf D}^\rho  \vert \phi ({\boldsymbol \xi};{\bf r})\rangle\vert \chi_{{\bf m} }({\boldsymbol \xi})\rangle.
\end{equation}
 as 
 \begin{equation}
 \label{as}
\langle n | {\bf D}^\rho | i \rangle = \langle\chi_{{\bf m}_{vc}}({\boldsymbol \xi})\vert \mu_{\bf{q}_c,\bf{q}_v}^\rho({\boldsymbol \xi})\vert \chi_{{\bf m} }({\boldsymbol \xi})\rangle.
\end{equation}
with 
\begin{equation}
\label{with}
\mu_{\bf{q}_c,\bf{q}_v}^\rho({\boldsymbol \xi})=\langle \phi_{\bf{q}_c}({\boldsymbol \xi};{\bf r})\vert\hat {\bf D}^\rho({\bf r}) \vert\phi_{\bf{q}_v}({\boldsymbol \xi};{\bf r})\rangle_{{\bf r}}.
\end{equation}
The matrix elements of the dipole operator ${\bf D}$ between two Born-Oppenheimer electronic states is the \textit{transition moment} $\mu_{\bf{q}_c,\bf{q}_v}^\rho({\boldsymbol \xi})$  connecting valence level ${\bf{q}_v}$ and conduction band electronic levels $\bf{q}_c$.
 $\mu_{\bf{q}_c,\bf{q}_v}^\rho({\boldsymbol \xi})$ is written for light  polarization $\rho$; the subscript ${\bf r}$ indicates that only the electron coordinates are integrated.  Note that 
$\mu_{\bf{q}_c,\bf{q}_v}^\rho({\boldsymbol \xi})$, if it does not vanish, is explicitly a function of phonon coordinates ${\boldsymbol \xi}$. 

 In terms of the transition moments, the Raman amplitude reads (using only the resonant term of the two)
 \begin{eqnarray}
 \label{KHD2}
  \alpha_{ {\bf m}_i,{\bf m}_f}^{\rho,\sigma} &=&\frac{1}{\hbar}   \sum_{v,c,{\bf m}}\left [  \frac{\langle  \chi_{{\bf m}_f}({\boldsymbol \xi}) | {\mu_{{\bf q}_v, {\bf q}_c}^{\sigma}({\boldsymbol \xi})}^{\dagger} | \chi_{{\bf m}_{vc}}({\boldsymbol \xi}) \rangle \langle \chi_{{\bf m}_{vc}}({\boldsymbol \xi}) | {\mu_{{\bf q}_c,{\bf q}_v}^{\rho}}({\boldsymbol \xi}) |\chi_{{\bf m}_i}({\boldsymbol \xi}) \rangle}{E_{{\bf m}_i} - E_{v,c,{\bf m}} + \hbar \omega_I - i \Gamma_{j,{\bf m}}}\ \right ] \nonumber \\
  &\equiv& \frac{1}{\hbar}   \sum_{v,c,{\bf m}}\left [  \frac{\langle   \varphi_{{\bf q}_v,{\bf q}_c,{\bf m}}({\boldsymbol \xi})  |   \chi_{{\bf m}_{vc}}({\boldsymbol \xi}) \rangle \langle \chi_{{\bf m}_{vc}}({\boldsymbol \xi}) | \varphi_{{\bf q}_c, {\bf q}_v,{\bf m}_i}({\boldsymbol \xi}) \rangle}{E_{{\bf m}_i} - E_{v,c,{\bf m}} + \hbar \omega_I - i \Gamma_{{\bf q}_j,{\bf m}}}\ \right ]
\end{eqnarray}
 The sum over the intermediate states with energy denominators is of course the energy Green function, i.e. the Fourier transform of the time Green function propagator. Although we do not write out the time dependent expression here (supplements, section~\ref{TDPT}), we see that \textit{the valence state phonon wave function $ |\chi_{{\bf m}_i}({\boldsymbol \xi}) \rangle$ arrives in the conduction band multiplied and thus modified by the transition moment ${\mu_{{\bf q}_j,{\bf q}}^{\rho}}({\boldsymbol \xi}) $, including its phonon coordinate dependence, before any time propagation takes place. Moreover, that propagation does not further change the phonon populations. }
 
 The product of the transition moment and the phonon wavefunction  is called $\varphi_{{\bf q}_j, {\bf q},{\bf m}_i}({\boldsymbol \xi})$ above. When $\varphi_{{\bf q}_j, {\bf q},{\bf m}_i}({\boldsymbol \xi})$ is re-expanded in all the phonon modes $\vert \chi_{\bf m}\rangle$ in all the conduction bands, some ${\bf m}$ will differ in the occupation numbers compared to the unmodified initial lattice occupation  ${\bf m}_i$ (recall  that because there is no geometry or force constant change, the phonons are the same in all valence and conduction bands). { \bf Every electron-hole pair production event  carries some amplitude for no phonon change, plus simultaneously a finite amplitude  for  phonon creation or destruction   relative to the initial phonon wave function.}
 
         The transition moment's  dependence on nuclear separation or equivalently  phonon displacements is unquestionable,   not only on direct physical grounds and explicit calculations, but also because there would be no  off-resonant Raman scattering without it. The derivative of the polarizability with nuclear coordinates, i.e. the Placzek polarizability, would vanish without   coordinate dependance of the transition moment.   If the transition moment is (unjustifyably) set to a constant independent of phonon coordinates, no Raman scattering occurs at all in KHD theory for graphene.
         
                         Due to dilution of the delocalized $\pi$ orbital amplitude over the infinite graphene sheet,  it is almost obvious that the Born-Oppenheimer potential energy surface is  unchanged after a  single electron-hole pair excitation from the ground electronic state.      No  new forces on the nuclei arise upon a single electron-hole pair formation, and no geometry changes take place. Thus the independence of $\vert \chi_{{\bf m}_{vc}}(\boldsymbol \xi)\rangle$ on the electron-hole state denoted by ${\bf q}_v{\bf q}_c$ (the occupations indicated by $vc$ may vary, but the phonon wavefunctions themselves do not depend on the electron-hole pair created).

The intermediate, typically conduction band (assuming no doping) Born-Oppenheimer states $ \vert\psi_{{\bf q}_v{\bf q}_c,{\bf m}}^{B.O.}({\boldsymbol \xi},{\bf r}) \rangle$ with energy $E_{v,c,{\bf m}}$ range over  resonant and non-resonant (very small or not so small denominators, respectively) states  including all  with nonvanishing matrix elements   in the numerator.   Only those  with  Pauli and momentum matched electron-hole pairs and electron-hole-phonon triplets can be nonvanishing; it is not necessary that they be ``resonant'', i.e. minimize the denominator, in order  to contribute importantly. Rather,     a range of states are collectively important, the range depending on $\Gamma$.  Although not relevant to graphene, even when none of the denominators reach resonance, there is still sufficient Raman amplitude to be quite visible as off resonance Raman scattering.  The final Born-Oppenheimer state $  \vert	\chi_{{\bf m}_f}(\boldsymbol \xi)\rangle$ with energy $E_{ {\bf m}_f}$   conserves total system plus incident and emitted photon energy, and may  differ from the initial state only by zero, one,  two, ...   phonons.  The electron has filled  the hole and the initial electronic state is restored.   
  
    An exception to this notation is the filling an empty level created by doping,  leaving the hole created by the photon unfilled.   All the electronic states reside exactly on the electronic band surface, here on the Dirac cone near the Dirac point.

It is good to keep in mind that the vast majority of   states $\vert\phi_{{\bf q}_v{\bf q}_c}({\boldsymbol \xi};{\bf r})\rangle\vert \chi_{{\bf m}_{vc}}({\boldsymbol \xi})\rangle$ (violating momentum conservation  with respect to the initial state for example)  in the KHD sum have vanishing matrix elements and do not lead to Raman (or even Rayleigh) scattering. Some terms   (or really a small group of terms - see discussion in section~\ref{KHDT} about backscattering ) give non-vanishing matrix elements and  if they include phonons, they are usually still incapable of Raman  emission. An example is a D phonon produced in absorption in a clean sample, unable to emit a Raman photon due to the Pauli blocking effect of the electron recoil.  Such states eventually relax and thermalize by a cascade of e-e and e-ph inelastic processes.	Transition 
 (see section~\ref{sliding}) makes this worse in that even elastic backscattering by defects does not  alleviate the Pauli blocking for the majority D phonons produced. These ``wasted'' phonons, not appearing in the Raman spectrum, are always  in the majority compared to ``successful'' phonons. It is not the phonons themselves that fail, but rather their associated, Pauli-blocked  conduction electrons.


Even if no phonon is produced at the time of absorption, the chances of achieving Rayleigh emission or phonon production upon emission are small, given the withering removal of electrons from   eligibility by inelastic e-e  scattering, raising the near certain specter  of Pauli blocking. This occurs on a femtosecond timescale (see \cite{coherence,lucchese2010quantifying,brida2013ultrafast,breusing2011ultrafast,HeinzUltrafast}).

\subsection{Phonon adjusted   electronic transitions}

A fraction of resonant   conduction band eigenstates    $\vert\phi_{{\bf q}_v{\bf q}_c}({\boldsymbol \xi};{\bf r})\rangle\vert \chi_{{\bf m}_{vc}}({\boldsymbol \xi})\rangle$ differ by one or more  phonons relative to the valence band state.  The energy of the Born-Oppenheimer eigenstates is a sum of phonon and electronic components,   \textit{so  the largest, resonant terms in the KHD sum equation~\ref{bopp} necessarily \textit{arrive} with  the electronic transition energy correspondingly lowered if a phonon is activated in the conduction band eigenstate relative to the valence state  or raised if a phonon is de-activated in the  conduction band eigenstate relative to the valence phonon wave function}. \textit {This  key fact, missing in DR,   alters electronic transition energies and phonon frequencies for dispersive bands. }  For phonon creation, the valence energy of the hole is raised, and the conduction band energy  lowered, keeping $q$ the same in both, thus   creating momentum matched (possible except needing elastic backscattering) holes and particles free of Pauli blocking.  We term this   pre-payment a ``diminished'' electronic transition  in the case of phonon creation (leading to  Stokes scattering), and an ``augmented'' electronic  transition   in the case of phonon annihilation (leading to anti-Stokes scattering).   It is important to note  (see also section~\ref{nvD}) that it is very difficult to produce an electron-hole triplet (i.e. with a phonon change) off-resonance in a mode requiring backscattering.  Thus the dispersive modes are more or less locked into a given diminishment or enhancement depending on the laser energy.

\begin{figure}[h] 
    \centering
    \includegraphics[width=6.5in]{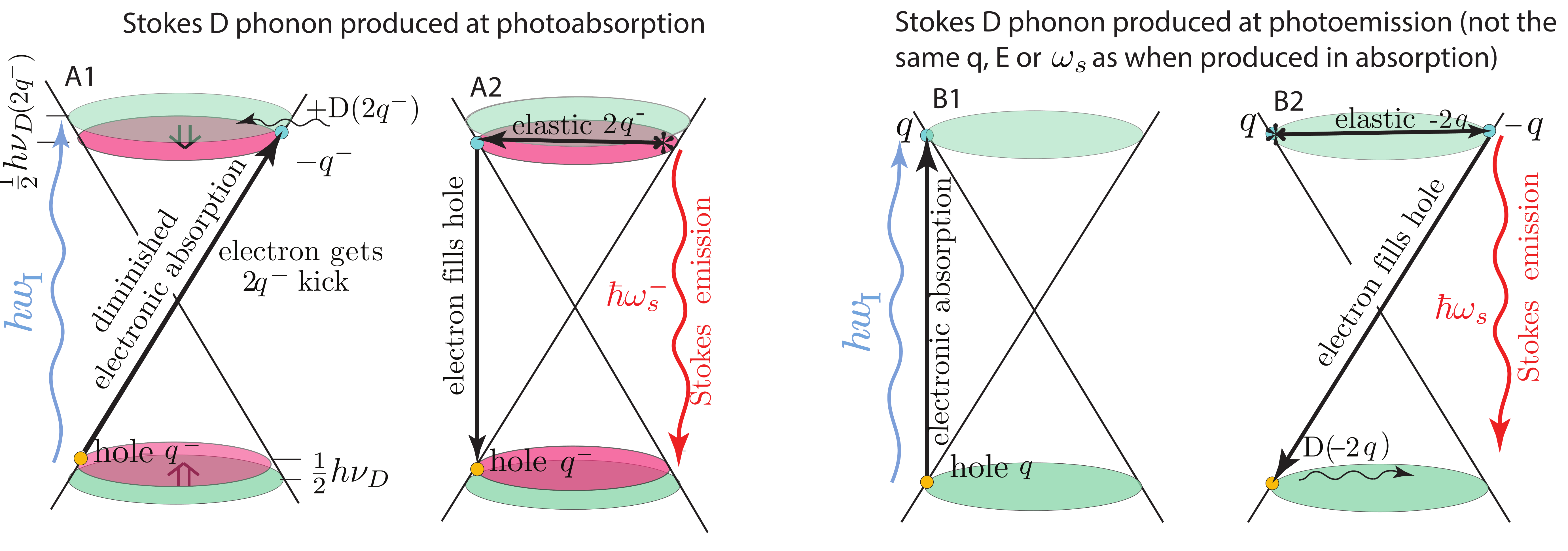} 
    \caption{ A.  Phonon creation in absorption, with diminished electronic energy.     A1:  The electron wave vector ${\bf q}$ is lowered to ${\bf q}^- $, affecting the  phonon energy through its dispersion.  The blue transition represents    photoabsorption with no phonon produced. $\nu_D$ is the effective frequency of the diminished transition.  A2. Defect elastic backscattering  allows matched electron-hole (no Pauli blocking) Raman Stokes emission. B. Stokes D band phonon creation in emission. B1. Full absorption of the photon by the electron-hole pair. B2. Defect induced elastic $2 {\bf q}$  backscattering of the conduction band electron,   allowing Raman Stokes emission  and D(2q) phonon production perfectly matched with the hole.    The energies of the D phonons produced in absorption and  emission differ, according to ${\bf q}^- $   \textit{versus}   ${\bf q}$ and the dispersion of the D band.  Anti-Stokes scattering (not shown) proceeds very similarly, except   augmentation rather than diminishment applies. 
    }
    \label{fig:actusl}
 \end{figure}

Near resonance, the pseudomomentum $ \bf{q^-}$ is determined by  the requirement that the energy cost of creating the electron-hole-phonon triplet  is just the laser photon energy. Typically, most of the energy needed is electronic, on the order of 1.5 or more eV with the phonon energy on the order of several 0.1 eV, which is not ignorable. The electronic transition energy is given by $E_e = E^c({\bf q^-}) -  E^v({\bf q^-})= h\nu_I -E_{phonon}(2\bf{q^-}) $, where superscripts $c, v$ refer to conduction and valence bands.   As $h\nu_I$ is changed, $\bf{q^-}$ of the phonon changes according to the valence and conduction band dispersion, which ideally is a Dirac cone structure with light-like linear dispersion of both bands. As  $\bf{q^-}$ changes, the phonon energy $E_{phonon}(2\bf{q^-}) $ changes according to the well known positive dispersion of about 50 cm$^{-1}$ per electron volt for the D band.  See figure~\ref{fig:actusl}, A. For simplicity, we  usually use   intravalley diagrams even when (in some cases) the process is intervalley.

If  in a single phonon transition, the   phonon is created or destroyed  at the time of  emission,   the electron-hole pair formation is not diminished or enhanced.  All the initial photon's energy goes into the electronic transition, as in DR, and no phonon change is yet present. The momenta of electron and hole are created in a matched pair (again, near resonance) with pseudomomentum ${\bf q}$. The conduction band electron may then elastically backscatter ${\bf q}\to \bf{-q}$, in the presence of defects. This allows recombination with the hole of momentum $\bf{q }$, provided it creates a phonon at $-2{\bf q}$ at that precise moment of emission (due to coordinate dependence of the transition moment).  The emitted light must be of the proper    Raman shifted frequency to conserve energy. The processes described are shown in figure~\ref{fig:actusl}, B. Since in this case  $\bf q$ did  not suffer diminishment in absorption, it will produce a phonon with slightly larger magnitude   $-2{\bf q}$ (no ``-'' on the  ${\bf q}$ is present) than if produced in absorption with its diminished transition.  The Raman shifts are thus slightly  different depending on whether the phonon was created at the time of emission or absorption, since given  dispersion the phonon will have a slightly different energy with a different  $\bf q$.

\textit{This means that all entries in phonon dispersion diagrams made  based on the DR model need a small correction, because   half of the electron-hole transitions are not the one DR supposes, differing instead by a phonon energy. } The correction is small, and contributes to the broadening (the addition of for example $\bf q$ and ${\bf q}^-$ peaks) as well as slight shift of the center frequency.

  The terms in the KHD sum    constructively and destructively interfere with one another before the  square is taken. The sum includes many different momentum conserving electron-hole pairs and electron-hole-phonon triplets leading to  the same  final states of the graphene system.   Processes leading to different final states (e.g. a different final phonon type or energy) appear in separate, non-interfering terms.

      \vskip .07in
  \noindent  \subsection{The  transition moment for graphene}
  \vskip .07in
 \label{gtm}

%
Consider the integral involving transition between a valence Bloch orbital at pseudomomentum ${\bf q}_v$,  described in terms of Wannier functions $\alpha_{{\bf q}_v}({\bf r}-\bf{R_{{\bf q}_v}}({\boldsymbol \xi})) $,  and a conduction Bloch  orbital at pseudomomentum ${\bf q}_c$, assuming only nearest neighbor ($\bf{A}$  with nearest $\bf{B}$)  interactions.  The transition moment becomes:  
\begin{eqnarray}
\mu_{{\bf q}_v{\bf q}_c}^\rho({\boldsymbol \xi})&=&\sum\limits_{\bf{A},\bf{B}}\int \! d{\bf r} \ e^{-i {\bf q}_v\cdot {\bf R}_{{\bf{A}}}({\boldsymbol \xi}) }\alpha_{\bf{A}}({\bf r}-\bf{R_{\bf{A}}}({\boldsymbol \xi}) ) \ \hat D^\rho\  e^{i {\bf q}_c\cdot {\bf R}_{\bf{B} }({\boldsymbol \xi}) }\alpha_{\bf{B}}({\bf r}-\bf{R_{{\bf{B}}}}({\boldsymbol \xi}) ) + c.c. 
\nonumber \\
&=& \sum\limits_{\bf{A}}e^{-i ({\bf q}_v-{\bf q}_c)\cdot \bf{R_{\bf{A}}}({\boldsymbol \xi}) }\sum\limits_{j=1}^3\   e^{ i{\bf q}_c\cdot \vec \delta_j({\boldsymbol \xi})}\ D_{\bf{A},\bf{B}_j}^\rho({\boldsymbol \xi})  \nonumber \\  &\equiv &  \sum\limits_{\bf{A}}e^{-i ({\bf q}_v-{\bf q}_c)\cdot \bf{R_{\bf{A}}}({\boldsymbol \xi}) } {\mathcal F}^\rho({\bf q}_c,\bf{A},{\boldsymbol \xi} )
\end{eqnarray}
$\vec \delta_j({\boldsymbol \xi})$ is a nearest neighbor vector, i.e. $\bf{B}_j= \bf{A} +\vec \delta_j ({\boldsymbol \xi})$.  ${\bf q}_v,{\bf q}_c$ are Bloch pseudomomenta in the valence and conduction band, respectively.  Here we have given only the simplest form of the transition moment; in fact in our calculations we use density functional theory modified Wannier wave fuctions as a function of nuclear coordinate; see section~\ref{tight}.

 The sum and therefore the transition moment vanishes at the equilibrium position of the lattice ${\boldsymbol \xi}_0$ unless ${\bf q}_v = {\bf q}_c$ or ${\bf q}_v - {\bf q}_c= \bf{K}$, a reciprocal lattice vector,  since ${\mathcal F}^\rho({\bf q}_c,\bf{A},{\boldsymbol \xi}_0)$ is  the same function of ${\bf q}_c$  for all $\bf{A}$. However, it is not the transition moment at a single configuration of the nuclei that is required, but rather the integral over phonon matrix elements, i.e. equation~\ref{as}. Suppose   ${\bf q}_v = {\bf q}_c$, so we are considering a $\Gamma$ point vibration. Clearly the exponential is unity.  But 
 it is easily seen that $\partial {\mathcal F}^\rho({\bf q}_c,{\bf A},{\boldsymbol \xi})/\partial { \xi}_G \ne 0$ at ${\boldsymbol \xi}  = {\boldsymbol \xi}_0$, i.e.  ${\mathcal F}^\rho({\bf q}_c,{ \bf A},{\boldsymbol \xi}_0)$ is odd at ${\boldsymbol \xi}_0$ about either G mode. (Moving the $ {\bf A}$ lattice up slightly and  the ${ \bf B}$ lattice down by the same amount is a distortion along a G mode  $\Gamma$ point vibration). Thus the transition moment can induce changes by one quantum in the G modes, at the   $\Gamma$ point.  
 
A phonon of pseudomomentum $ \bf{k}={\bf q}_v-{\bf q}_c$ can be induced by  lattice distortion ${\boldsymbol \xi}_k \ne {\boldsymbol \xi}_0$ and the transition moment if 
$
\mu_{{\bf q}_v{\bf q}_c}^\rho({\boldsymbol \xi}_k) =\sum_{\bf{A}}e^{-i ({\bf q}_v-{\bf q}_c)\cdot \bf{R_{\bf{A}}}({\boldsymbol \xi}_k)} {\mathcal F}^\rho({\bf q}_c,\bf{A},{\boldsymbol \xi}_k)$
becomes   non-vanishing for   ${\bf q}_v - {\bf q}_c = \bf{k}$ or ${\bf q}_v - {\bf q}_c= \bf{k}\pm {\bf{K}}.$ This happens due to    periodic undulations in ${\mathcal F}^\rho({\bf q}_c,\bf{A},{\boldsymbol \xi}_k)$ arising from  displacement of the lattice $\bf{R_{\bf{A}}({\boldsymbol \xi}_k)}$ and $\bf{R_{\bf{B}}({\boldsymbol \xi}_k)}$ according to a phonon with wave vector 
 $\bf{k}$.  However, unless strict conditions are met, such phonons, though present, are associated with a conduction band electronic level that is Pauli blocked, keeping the phonon silent in the Raman spectrum.  Elastic backscattering  will not help, except for special cases, as described next. 
   If hole doping is present, matters can be changed  in a fascinating way; see section~\ref{evidence},  below. 
   
 \subsection{Avoiding Pauli blocking}

 An  important pathway that avoids  Pauli blocking begins with   ${\bf q}_c^-=-{\bf q}_v^-$,  i.e.  $ \bf{k} = 2 {\bf q}_v^-$, where the superscript ``-'' refers to a diminishment of the electronic transition energy by the simultaneous creation of a $2 {\bf q}^-$ phonon (in the case of Stokes scattering); see below.  The electronic transition will be accordingly diminished by the energy of the $ \bf{k} =  2{\bf q}^-$ phonon.   The conduction electron suffers a ${\bf q}_v^- \to -{\bf q}_v^-$ kick to conserve momentum, and is Pauli blocked.  However in this case  subsequent defect elastic backscattering can re-align it with the hole,  allowing recombination. A photon is emitted at a frequency revealing the diminished Stokes shift of the $2 {\bf q}^-$ phonon production.    Since the diminished  ${\bf q}^-$ depends also on the incoming laser energy, the phonon dispersion will be revealed by changing that energy. A $2 {\bf q}^-$ modulation of the G mode for example will become Raman allowed by this mechanism.  The modest dispersion of that mode will give rise to a sideband to G, which we call G$^\prime$, visible only in the presence of defects.  This is exactly what is observed, see figure~\ref{fig:tracehilke}.
 
 Another   important case is ${\bf q}_v^-={\bf q}_c^-$,  i.e.  $ \bf{k} = 0$, which was discussed in section~\ref{gtm}.    This corresponds   to creating a $\Gamma$ point G phonon that carries no momentum,   but it still carries  energy.     Some of the photon's energy is  channeled directly into the G phonon energy, diminishing the (in the case of Stokes scattering) electron-hole transition energy, including shifting the $\vert \bf q\vert $ of the transition nearer to the Dirac cone $K$ point,     \textit{as if} lower energy light had been used: $0.185$ eV lower for a 1500 $\textrm{cm}^{-1}$ phonon. Raman emission is active because the electron and hole are born matched in ${\bf q}$ and ready to recombine. Fast   e-e scattering is the enemy of Raman emission, since if it occurs, the change in electron momentum  makes Pauli blocking  a near certainty.  (Studies point to a timescale of a  few  femtoseconds before irreversible relaxation of conduction band  electrons by e-e scattering. 
  Definitive experimental results\cite{coherence,lucchese2010quantifying,brida2013ultrafast,breusing2011ultrafast,HeinzUltrafast} affirm the extremely rapid  relaxation of   photo excited electrons due to e-e scattering.) The G band is indeed bright, and benefits too from off resonant contributions (unlike D), but the fact that a normally weak overtone, 2D, is  in fact perhaps 10 times brighter than G follows from a fascinating process; see    sections~\ref{sliding} and \ref{evidence} below  on sliding transitions.

      \vskip .07in
\section{Tight Binding and Density Functional Realization of Graphene KHD}
\label{tight}
  \vskip .07in
  \begin{figure}[h] 
   \includegraphics[width=6.5in]{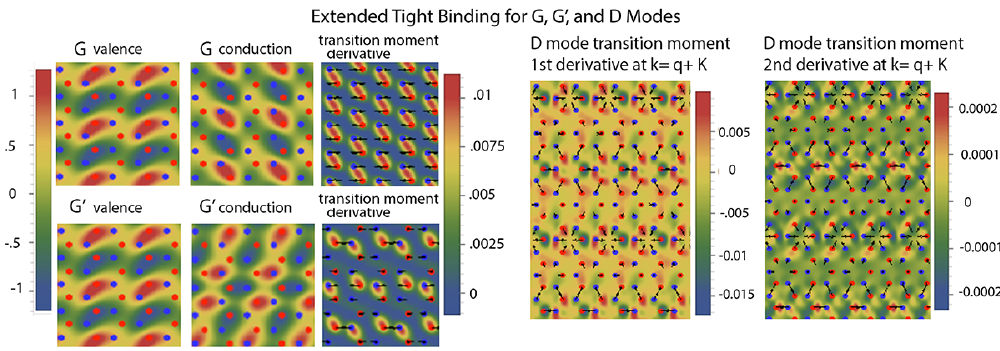} 
   \caption{(above) The extended tight binding results for the G and G$^\prime$ modes. The valence state is at the left.  The middle panel depicts the conduction band state in each case, and at the top right is shown the derivative of the transition moment with respect to the ${\bf k}=0$ G phonon coordinate, and the ${ \bf k}=2q$ G$^\prime$ (formerly  called D$^\prime$; see below)  phonon coordinate (bottom  right).  (below)  The first derivative of the transition moment in the D sideband mode direction  at ${ \bf K+q}$, ${ \bf K}=2\pi (\frac{1}{3 a}, \frac{1}{3 \sqrt{3} a}), {\bf q}=2\pi (0, \frac{1}{12 \sqrt{3} a})$. The transition moments  and their derivatives are integrals over the data in these figures.  (left, bottom) The first   derivative of the transition moment has amplitude of 1.2 in arbitrary units. (right, bottom) The  second derivative of the transition moment, which has amplitude 0.01. }
   \label{fig:gall}
\end{figure}

The simplest model Hamiltonian for the single-layer graphene involves only nearest-neighbor interactions. However, in the presence of   crystal distortions, the  hopping strength should vary with the pair distance. This is especially important to incorporate here, since this will contribute to the dependence of transition moments on geometry changes, needed in our KHD theory.
 
  A two-step {\it ab-initio} procedure is used to model realistic hopping parameters. First, the density functional theory (DFT) calculations are performed using the Vienna  \textit{ab initio} Simulation Package (VASP)\cite{vasp1,vasp2} with the exchange-correlation energy of electrons treated within the generalized gradient approximation (GGA) as parametrized by Perdew, Burke, and Ernzerhof (PBE)\cite{pbe}. To model the single layer graphene, a slab geometry is employed with a 20 \AA $ $ spacing between periodic images to minimize the interaction between slabs, a $450$ eV cutoff for the plane-wave basis and a reciprocal space grid of size 19 $\times$ 19 $\times$ 1 for the 1 $\times$ 1 unit cell. 

Based on the DFT band structure and Bloch waves, the Kohn-Sham Hamiltonian can be transformed into a basis of maximally-localized Wannier functions (MLWF)\cite{mlwf} using the Wannier90 code. The initial projections for Wannier functions are the atomic   \textit{p}$_z$ orbitals and the transformed Hamiltonian is the \textit{  ab-initio} tight-binding Hamiltonian. By varying the positions of the basis atoms, the hopping strength $t$ for different pair distances $r$ can be extracted and its empirical formula at the linear order reads:
\begin{equation}
t(r) = t(r_0) +f_1 (r-r_0)
\end{equation} where $r_0$ = 1.42 \AA, $t(r_0)$ = -2.808 eV and $f_1$ = 5.058 eV/\AA.
 Figure~\ref{fig:gall} gives the extended tight binding results for the G and G$^\prime$ modes   at the top,  in two three-panel strips.     The middle panels of each strip depict  the conduction band states, and the rightmost  panels  show the derivative of the transition moment with respect to the $\bf{k}=0$ G and the   $\bf{k}=2q$ G$^\prime$ phonon coordinate,  shown by the small arrows within the images, (notice the undulations in the atomic displacements).
 The bottom two panels reveal first and second derivatives of the transition moments for the D mode at  ${ \bf K+q}$, where ${ \bf K}=2\pi (\frac{1}{3 a}, \frac{1}{3 \sqrt{3} a}), {\bf q}=2\pi (0, \frac{1}{12 \sqrt{3} a})$. 
 
 Consider a transition from  $K+q^-$ (valence) to  $K-q^-$ (conduction), with   $K =2\pi*(\frac{1}{3 a}, \frac{1}{3 \sqrt{3} a}),  q^- =2\pi (0, \frac{1}{12 \sqrt{3} a})$. $  K$  is exactly at the Dirac cone,  $  q$ gives a small displacement from the cone center, $a$ is the carbon-carbon  bond length at equilibrium. In this case, the electron has a momentum change of $  -2q^- $, and the phonon has momentum $  2q^-$. (For comparison, using  the graphene sheet  only partly depicted  in figure~\ref{fig:gall}, the constant part of the transition moment  has amplitude about 60 in arbitrary units). With displaced  atoms of amplitude 0.01 \AA \   (adjusted according to $ 2q^-$ modulation), the first   derivative of the transition moment has amplitude of 1.2; the second  derivative has amplitude 0.01. 
 
 We find that the electronic transition moment has a robust first derivative along any choice of the  independent and degenerate G mode phonon coordinates, and this accounts for their presence in the Raman spectrum.  Figure~\ref{fig:gall} 
shows the local transition moment along one of those choices.  It is seen to be perfectly repetitive with the unit cell translation vector    as befits a $k=0$ optical mode. Integration of this 2D local transition moment over space yields the transition moment at the given nuclear positions, and repeating this after a phonon distortion reveals the phonon coordinate dependence of the transition moment.

Thus D should be robust (with sufficient elastic backscattering), but simultaneous production of 2D seems in doubt.  Since 2D is the brightest band, an explanation is needed, and is given below in    sections~\ref{sliding} and \ref{evidence}.
 
     \vskip .07in
\section{Analysis of Graphene Raman Band Structure with KHD}
  \vskip .07in
  
\begin{figure}[htbp] 
   \centering
   \includegraphics[width=6.6in]{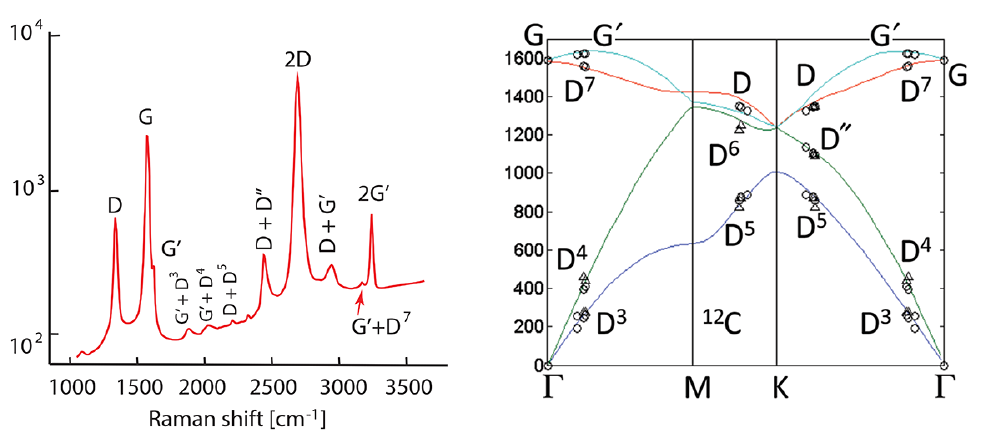} 
   \caption{(left) Averaged defective  graphene $^{12}$C Raman spectrum on a Si substrate, redrawn from reference \cite{hilke}.  (right) Placement of phonon dispersion points found and assigned through Raman spectroscopy  by Hilke   \textit{et. al.} \cite{hilke}, reproduced with permission. D$^{\prime}$ symbol replaced by G$^{\prime}$, appropriate to a sideband of G. }
   \label{fig:tracehilke}
\end{figure}

  We now go though a few prominent characteristics of graphene Raman spectrum discovered and discussed over the years, explained (very directly) by the KHD theory.
  In figure~\ref{fig:tracehilke} a Raman spectrum obtained by the Hilke group is an average of 60 different  samples, each with defects,   in order to bring out weak bands forbidden in clean, perfect graphene crystals.  The D band is one such case, while 2D is allowed and  bright in pure samples.   
\vskip .07in\subsection{ Origin of G band intensity} \vskip .07in
 The  constant part of the electronic  transition moment for arbitrary electronic $\bf q$ is non-vanishing and responsible for most of  the light absorption in graphene. It delivers electrons to the conduction band without creating a phonon.  A  phonon's creation (or annihilation) is the  result of a changing electronic transition moment as its  coordinate is displaced  from equilibrium.   The more rapid the change in transition moment as a function of phonon coordinate, the more likely is the phonon's creation.

%

\begin{figure}[htbp] 
   \centering
   \includegraphics[width=4in]{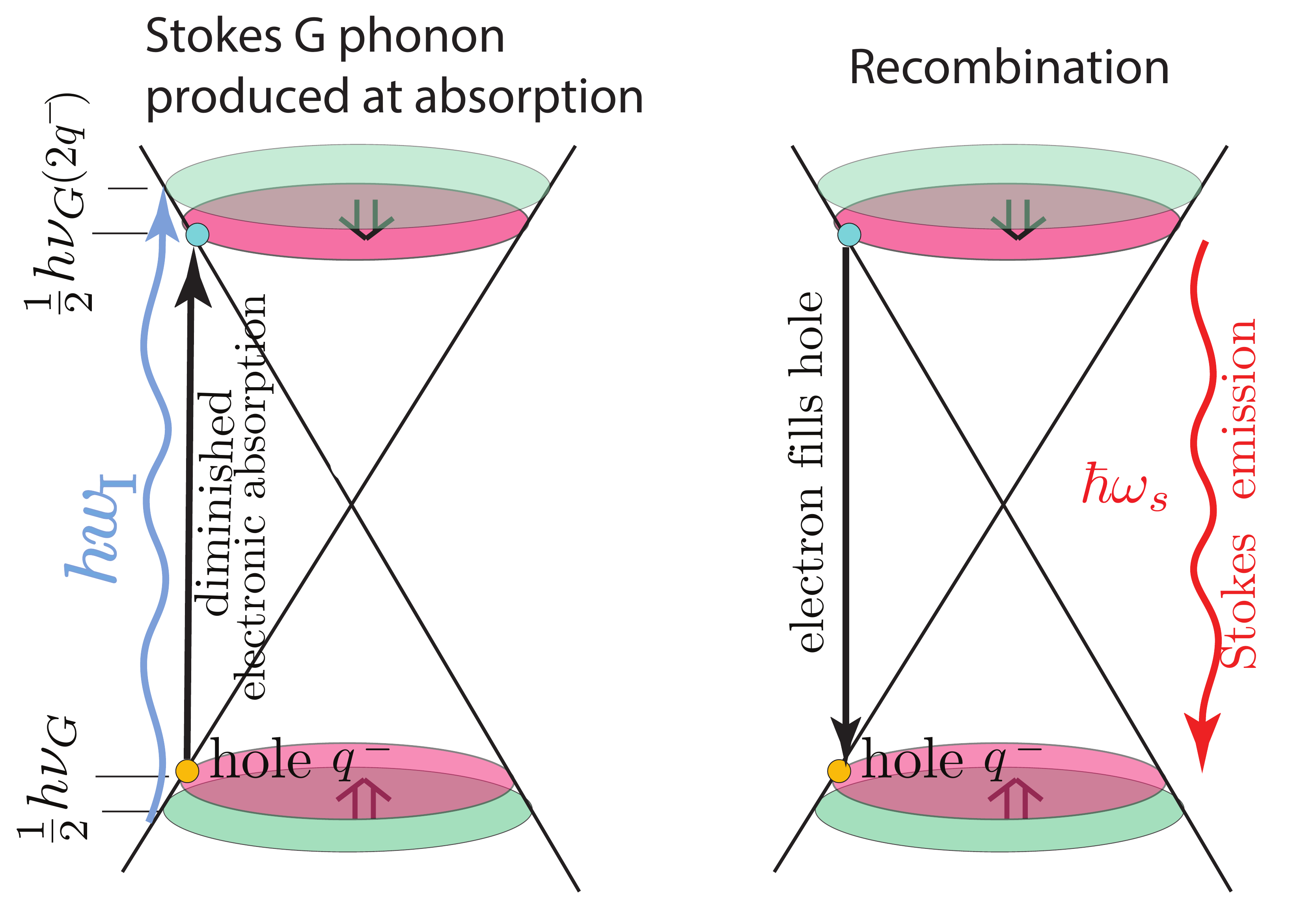} 
   \caption{Production of a G phonon in absorption, according to KHD extended to graphene.}
   \label{fig:gmode}
\end{figure}

The G modes   have no dispersion;  the same ${\bf q}=0$ mode is produced independent of laser frequency. 
The G mode may also be produced in  emission.  The KHD expression has the same transition moment promoting either event.  The production of the G by either means is a small minority of events in any case,  as is any phonon producing a successful Raman emission.  (It is often stated that typically only about 1 in $10^{11}$ or $10^{12}$  incident photons causes a Raman emission).
The mechanism based on KHD for G mode production in absorption is given in figure~\ref{fig:gmode}.

In section~\ref{bright}, we discuss the brightening of the G mode band due to hole doping and off resonant scattering. For the G-mode, the phonon production off resonance  occurs by the same mechanisms familiar in KHD that populate vibrational modes in smaller systems off resonance.

\subsection{No virtual process for   D mode production or others requiring backscattering}
\label{nvD}

The G mode does not require backscattering to be produced.  The electron is ready to fill the hole the moment in appears.  Not so the D mode. In order to produce a D, a recoil of the conduction electron is required.  Virtual processes are short lived; the electron must fill the hole quickly, leaving insufficient time for elastic backscattering.  The off-resonant electron and hole cannot remain, since  after some time the energy imparted becomes more certain; off-resonance, the energy is not sufficient to support this. 

The difficulty is seen in the equation~\ref{KHD2}.  If an off resonance state is created with a phonon and ${\bf q}_v=-{\bf q}_c$, without backscattering this situation arrives Pauli blocked at the second transition moment operator, which cannot right the situation without creating another, counter-propagating D phonon. Otherwise the numerator vanishes and we may say the virtual process does not exist unless the second, counter-propagating D phonon is produced.  This 2D scenario may indeed take place,   over wide  zones on the  Dirac cones, even for the same photon.  However    among these is a resonant D production and subsequent second D emission; the  resonant process is  expected dominate. Even the purely resonant  2D production for fixed phonon wave vector can slide up and down the Dirac cone, at a single photon frequency, thanks to the linearity of he dispersion. This is our first glimpse of the the  transition  sliding mechanism presented in section~\ref{sliding}
 \vskip .07in
 \subsection{Absence of 2G overtone} 
 \label{2Gabsent}
  \vskip .07in 
  Given the robust strength of the G band, at least a small overtone at 2G would be expected, yet the 2G band did not  make any appearance in the spectra of Hilke  \textit{et. al.} \cite{hilke}, where other  weak bands were seen for the first time.   We have been touting the uncommon  strength of the overtones in graphene, so this absence must be regarded as one of its mysteries and must be explained. There is no group theoretical ban on its existence, and like G it is  a $\Gamma$ point  mode, requiring no backscattering.   Why is it missing?

DFT-tight binding calculations of the transition moment  show that  the second derivative along the G phonon mode is about 2 orders of magnitude smaller than the first derivative. Since the intensity goes as the square of the transition moment, this  would wash out 2G if it were the only scenario for a 2G.    
 
However there is another possibility to consider: The robust linear slope in the transition moment  along the G mode could  be used twice, once in absorption and the second time in emission. Although not  a simultaneous production of the phonons, two G mode phonons will have been produced, and a Raman band would appear at 2G. The phonons would both be pseudomomentum $k=0$.   
It is easy to see however that the intensity for this process must  be extremely low: If there is an  amplitude of $0.025$ (probably much too high an estimate, but conservative for this purpose) for producing a single G mode in absorption, the amplitude for two G phonons, one in absorption and one in  emission, is $0.000625$.  This  corresponds to  a probability of two G phonons produced this way some 1600 times smaller than  the probability of a single G phonon production.   ``One up, one down'' does not lead to a visible signal.

Paradoxically, we will see  that this ``one up, one down''   mechanism \textit{ is the key to   the brightness  of the  2D band}, and other overtone bands.  The 2G mode is unable to increase its brightness by one up, one down ``transition sliding'' (section~\ref{sliding}),  because sliding requires a $k = 2q$ phonon be produced.  Sliding greatly amplifies the chance of producing a phonon,  for example in the case of 2D, or 2G$^\prime$ (formerly called 2D$^\prime$) since many simultaneous amplitudes are summed. Transition sliding is a key principle made possible by linear Dirac cone dispersion (section~\ref{sliding}), and a key Raman mechanism revealed here.
 
 \vskip .07in \subsection{The  G$^\prime$  [former D$^\prime$] and 2G$^\prime$ [former  2D$^\prime$] Bands }   \vskip .07in
 
 Nearby the $\Gamma$ point ${ \bf k}= 0 $  phonon, the transition moment can also give rise to   ${ \bf k}= 2{ \bf q^-}$  phonons, giving a momentum kick $2 { \bf q^-}$ to the conduction band electron. Elastic backscattering     makes recombination possible, revealing the existence of the sideband to G known as D$^\prime$.  As Ferrari  and Basko   suggested \cite{ferrariNat}, G$^\prime$ would be a better name for D$^\prime$.   We adopt this notation, in spite of the checkered history of the G$^\prime$ nomenclature, which used to denote 2D not many years ago.  
 
 The healthy 2G$^\prime$ band is the  overtone of G$^\prime$   and    does not require defect elastic backscattering.   Graphene Raman spectra in the literature are often cut off before its ca. 3200 cm$^{-1}$ displacement.  Its  frequency  is  close to twice that of G$^\prime$.     It seems the 2G$^\prime$  band owes its unexpectedly large intensity in the absence of defects to the same transition sliding mechanism that benefits 2D; see the next section.


%

                     \begin{figure}[t] 
                         \centering
                         \includegraphics[width=6.5in]{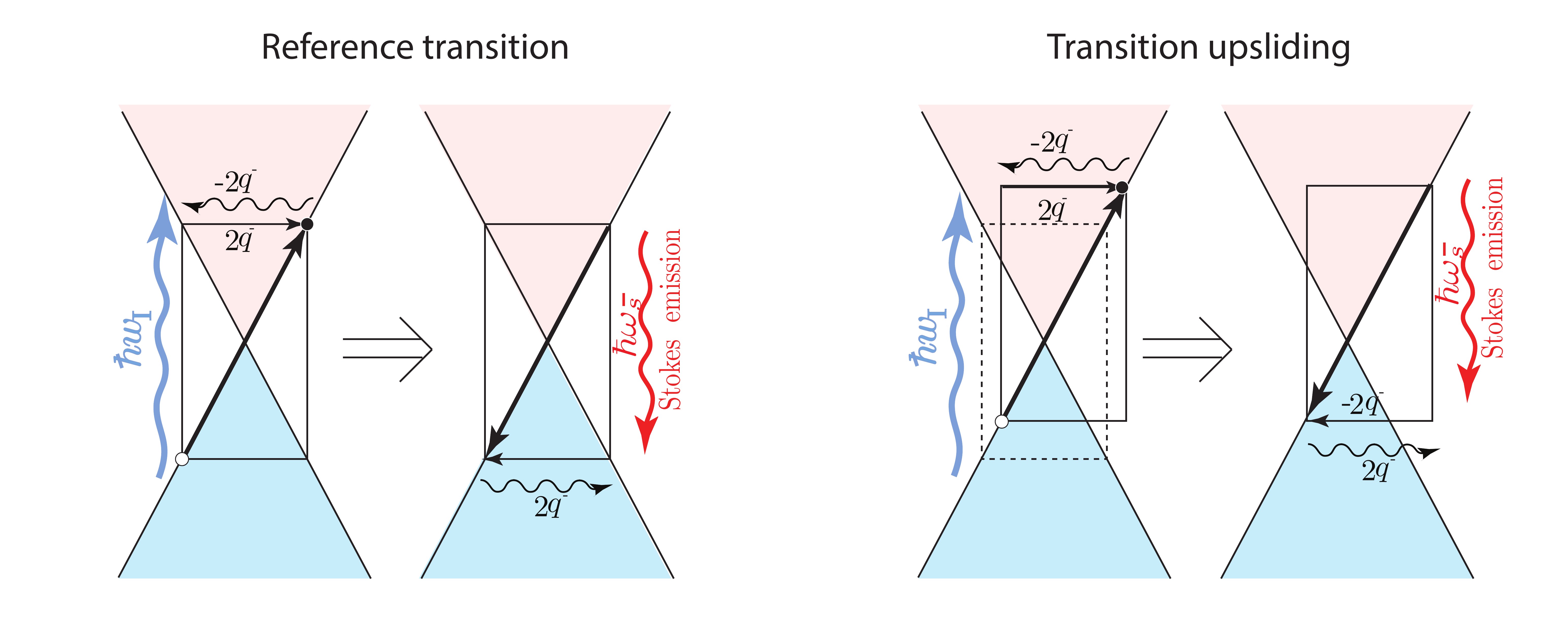} 
                         \caption{ \textbf{Transition sliding on  a Dirac cone.}  
                   (left) The nominal, symmetric  ``Double D'' (2D) transition with   energies equal above and below the Dirac point. Absorption is followed, without any backscattering, by emission along the reverse path, producing a second phonon of opposite momentum. (right)  By moving the rectangle up, not changing any dimension, we find  it still fits  absorption and emission  transitions  (it could also slide downward) along the Dirac cone. The reference transition is shown as a dashed rectangle. The phonon $2q$ is unchanged, since $q$ is  sas is the electronic transition energy and incident photon energy.                                                                                                                                                                                                                              }
\label{fig:sliding2}
\end{figure}    

 \vskip .07in
 \section{Sliding Phonon Production}
  \vskip .07in
  \label{sliding}
  
  It is a consequence of linear Dirac cone dispersion that the transition making a D phonon in absorption, shown at the left in figure~\ref{fig:sliding2}, is allowed over a range of valence to conduction electron-hole transitions all for the same incident laser frequency and all  producing the same phonon.  The therefore result in the same final state and contribute  to the same Rama amplitude, which is  squared to give the cross section.
  
The brightness of the 2D overtone mode (and of other $k=2q$ overtone bands like 2G$^\prime$) is a consequence of these continuously many simultaneous transition amplitudes, each one starting from a different valence band level, all for the same incident laser energy and all producing the same D phonon in absorption and one of opposite pseudomomentum in emission. Amplitude for electron-hole and electron-hole-phonon production appears simultaneously up and down the Dirac cone, vastly extending   the nominal  symmetrical $K-q$ to $K+q$ transition (giving a -2q kick to a new phonon).  That is, a  continuum of  $K-q +\delta q$ to $K+q+ \delta q$ transitions  is available, all generating the same    $-2q $ phonon.  This is possible due to   the linear dispersion of  the  Dirac cone.  Since the $2q$ phonons are the same, the sliding transition amplitudes all share the initial state $\vert i\rangle$ and final state $\vert f\rangle$ and add coherently, before the square is taken.  The energy $E_{v,c,\bf{m}}$ in the sum over electron-hole-phonon triplets $\vert\phi_{{\bf q}_v{\bf q}_c}({\boldsymbol \xi};{\bf r})\rangle\vert \chi_{{\bf m}_{vc}}({\boldsymbol \xi})\rangle$  is not changed by sliding: the same photon is absorbed, and always put into making the same $-2q_v$ phonon with the balance put  into the electronic part of the transition;  $E_{v,c,\bf{m}}$   is highly degenerate.  This is true of every distinct $E_{v,c,\bf{m}}$, resonant or not: a continuum of resonant and off-resonant states $\vert\phi_{{\bf q}_v{\bf q}_c}({\boldsymbol \xi};{\bf r})\rangle\vert \chi_{{\bf m}_{vc}}({\boldsymbol \xi})\rangle$ with nonvanishing transition moments ranges over  valence holes and conduction electrons (plus $2q$  phonons)  with   energy near $E_i + \hbar \omega_I$.

The sliding transitions are already built into the KHD equation~\ref{bopp}; one has only to examine the possible intermediate states  to see that they are present. Their ultimate limitations  will be a rich topic for future study: trigonal warping and distortion away from linear dispersion at energies farther from the Dirac point will cause slightly different energy phonons to be produced; these will add to the intensity non-coherently, only after the square.  They will broaden the transition but are intrinsically less bright due to diminishing  constructive interference.  Also limiting the sliding is the density of states, diminishing to 0 as the band edges are reached. 

We have checked the transition moments for sliding  transitions with our tight binding model.  For example, the transition from an occupied conduction band level at $K+q$ to an empty conduction band at $K+3q$.  The transition moment  first derivative is not significantly smaller than a transition from an occupied valence band at $K-q$ to an empty conduction band at $K+q$.  
 
 Sliding adds to the transition amplitude for production of  the first $2q $ phonon, but electrons produced by such sliding transitions   do not match the  holes they leave behind and are thus Pauli blocked in emission.  Elastic backscattering does not rescue these phonons from obscurity;  ($K+q +\delta q$, backscattered elastically, gives   $K-q -\delta q$, which does not match the original hole at  $K-q +\delta q$).    Below, we  note that  hole doping creates empty  valence bands, that \textit{can}, if conditions are right, accept emission from such conduction band states,  giving rise to a spectacular broadband ``electronic Raman'' emission as seen in reference \cite{chenCrommie}; see section~\ref{evidence}. (These authors provided  a   different interpretation involving ``hotband'' emission following electron-phonon scattering).
 \subsection{Reversing the path}
 The Raman-silent single phonon process becomes Raman active when the conduction band electron, produced by a sliding transition along with  a $k=2q$    phonon,  emits (without first backscattering) to the valence band \textit{along the  reverse path used in absorption}, (see figure~\ref{fig:sliding2}). A second,  oppositely propagating  phonon is released.  The electron is automatically  matched to the hole and recombines, without Pauli blocking of any sort, and in the case of D phonons a proper  matched 2D phonon pair has been produced. The cumulative effect of constructively interfering contributions  over a continuous sliding range  of the terms in the KHD sum (since the numerator of every term in the sum is an absolute value squared, before the overall square is taken) causes a large enhancement of the 2D Raman band intensity.     In graphite whiskers, the  intensity of the 2D overtone is found to be about 10 times stronger than the single phonon G mode, normally expected to be much stronger than an overtone band \cite{tan2001polarization}. As mentioned above, the G mode is not amenable to sliding transitions.

  Figure~\ref{fig:sliding2} shows the sliding scenario.   A large range of sliding  $\Delta E^v=\Delta E^c$, i.e. equal shifts in valence and conduction bands,  are available; all of these are resonant, not virtual,  transitions.   Compared to teh reference, non-sliding symmetric case, in up-sliding the valence wave vector has been shortened by  $\delta  q$, and the conduction wave vector has been lengthened by the same amount, so it remains a 2q$^-$ transition and giving a 2q$^-$ phonon production, just  as   when  $\Delta E^v=\Delta E^c=0$.  All the sliding transitions are independent amplitudes at the same photon energy simultaneously present,  and together they vastly enhance the probability of producing a pair of D phonons. 
  The density of states for both the initial and final electronic states will have a major effect on the propensity to slide various amounts.   

   The sliding mechanism explains many known facts of Raman scattering in graphene.  First and foremost, the brightness of the 2D and other overtone bands (mixed transitions can occur as long as the pseudomomentum is the same in absorption and emission) results from addition of a continuous range of sliding amplitudes, before the square is taken to give Raman intensity.  Second, the fact that the 2D peak strongly decreases with increasing doping or disorder is now explained easily: Doping can provide  much faster phonon-less emission pathways to   empty valence states that will diminish production of the emission producing another D phonon.   Elastic scattering of  the conduction electron (it does not have to be backscattering) by defects will also quench the  D 	emission probability by Pauli blocking and thus quench the 2D intensity.

     \vskip .07in
  \vskip .07in \subsection{The D band}  \vskip .07in

 As we have seen in figure~\ref{fig:sliding2}, the  electronic transition can slide up and down the Dirac cones and still produce a $2q^-$ D phonon.  The conduction band electron arrives at $q^- + \delta q$, having  shifted momentum by $\delta q\ne 0$ and getting a $2q^-$ kick from the phonon. It will   not match its hole   at  $-q^- + \delta q$,  even with elastic backscattering, since it would have momentum   $-q^- - \delta q$. Thus D phonon  production \textit {visible in the Raman signal} does not benefit from sliding.  The D phonons are present nonetheless; the electron is hung up in the conduction band and must relax by other means than Raman emission.
 

\vskip .07in \subsection{ D band  Stokes, anti-Stokes anomaly} \vskip .07in
 
The Stokes versus anti-Stokes frequencies in the D and 2D bands are   graphene Raman anomalies, discussed   first in pyrolytic carbon \cite{dressNoPerm,ferrariNat} and graphite whiskers \cite{tan2001}. There are two striking experimental results to explain:  (1) A difference between D band  Stokes and anti-Stokes frequencies.  In a small molecule, the Stokes and anti-Stokes  bands measure the same vibrational state, and there cannot be any difference between the two: they are symmetrically spaced across the Rayleigh line, i.e. 0 asymmetry.   For the graphene D band, the asymmetry  is instead about 8 or 9 cm$^{-1}$. (2) As shown in the next section, the 2D band  Stokes,   anti-Stokes asymmetry  is not twice the D band asymmetry, which would be expected because two  D phonons are produced, but close to 4 times the D band shift, or 34 cm$^{-1}$.  These numbers  emerge  simply from our KHD theory, without invoking virtual processes.  
 
\begin{figure}[h] 
   \centering
   \includegraphics[width=6.5in]{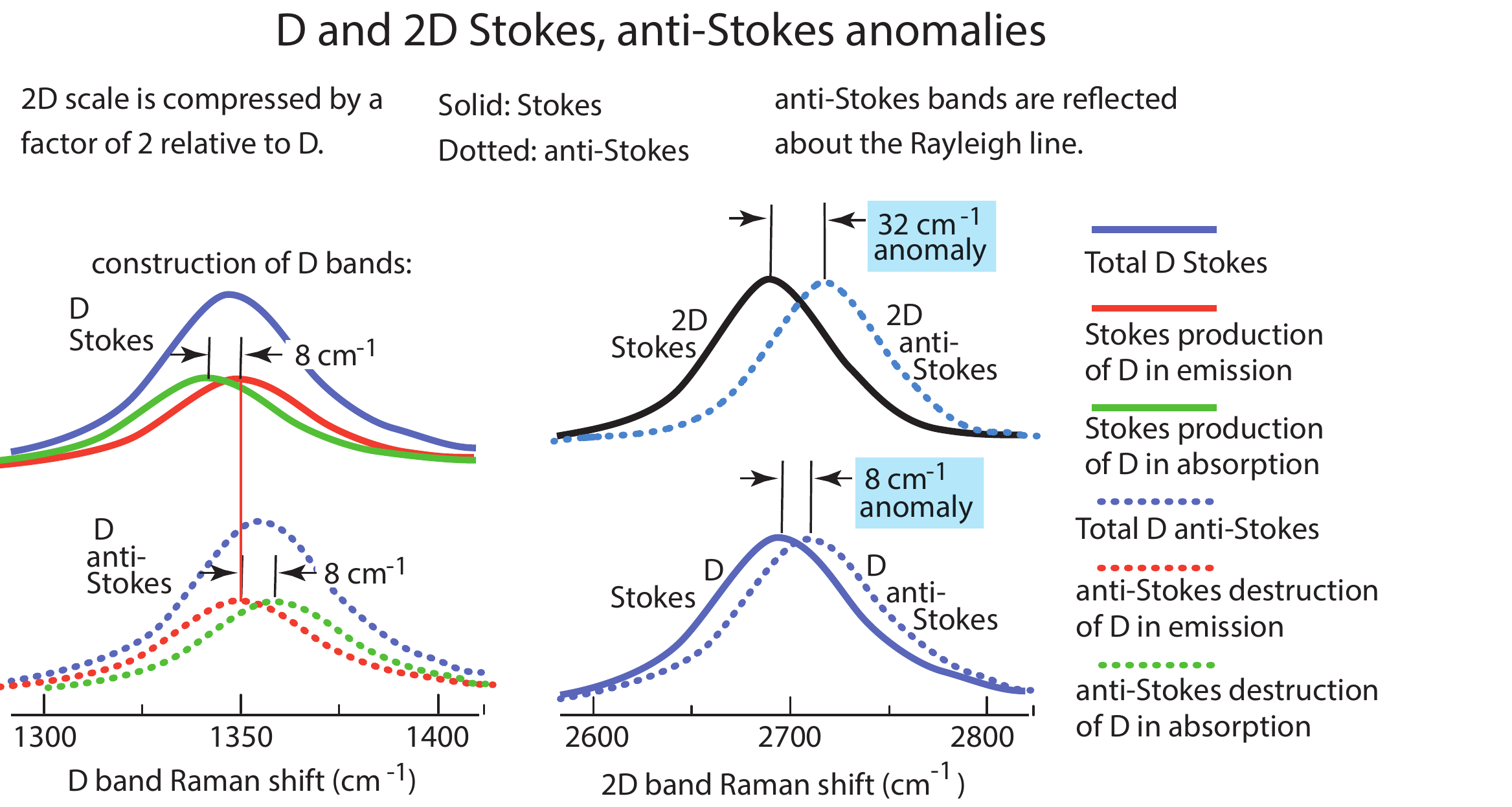} 
\caption{ The  prediction of Stokes and anti-Stokes Raman spectra of the D and 2D band positions and their anomalies,    in KHD theory.   The horizontal scale of the 2D-band spectra (top) spans twice the range of the D scale. Anti-Stokes processes  were brought across the origin of the Rayleigh line and are shown dotted. See text for details.}
   \label{fig:dressNoPerm}
\end{figure}

 The  D mode Stokes band is an average of    emission and absorption production, with emission production unshifted.  Production in absorption is shifted  down 8 cm$^{-1}$, by the dispersion of D and the diminishment cause by energy conservaton.  The average of the two is    4 cm$^{-1}$   \textit{closer} to the Rayleigh line than the undiminished emission production alone would be (see figure~\ref{fig:dressNoPerm}).    Similarly, the D mode anti-Stokes production in absorption  also consists of two bands,  overall 4 cm$^{-1}$ higher  in energy and   \textit{farther} from the Rayleigh band. At  3.5 eV and  1350 cm$^{-1}$ the Stokes vs. the anti-Stokes D phonons (reflected  about the Rayleigh line for comparison)  will differ by about 8.4 cm$^{-1}$. This is an anomalous  Stokes,  anti-Stokes  asymmetry  of about 8.4  wave numbers, in excellent agreement with experiment. Thus the anomaly has a simple explanation in terms of phonon production in absorption vs. emission.

 \vskip .07in \subsection{The 2D band} \vskip .07in
 
The 2D overtone band in pure graphene is the strongest line in the spectrum, even much stronger than the fundamental G band. The Kohn anomaly has been proposed as a contributor to the strength of   2D, and indeed it may be, but then G is weaker, also born at a Kohn anomaly, and  is a fundamental, not a normally weak overtone.  

It is important  to note that if two counter-propagating  D  phonons were actually produced  simultaneously in absorption,   there would be a doubling (two phonons) of a double diminishment of the electronic energy (since twice the energy is needed from the photon to produce both phonons at once).  This implies  a shift of 32 cm$^{-1}$ relative to the presumably equally important simultaneous emission production of two counter-propagating phonons, a transition that is not diminished in energy or $q$. This would imply that the 2D band would either be double  or a single peak  considerably broader than 32 cm$^{-1}$. This is not consistent with  experiments revealing  a  slightly asymmetric  line about 25 cm$^{-1}$ FWHM \cite{Heinz2Dshape}. 

But there is another possibility, just described,  that of producing a  D phonon in absorption and another in a mirror image emission. There are a continuum of such amplitudes which can also slide up and down the cone, part of the KHD sum, each contributing an imaginary part with the same sign to the total 2D amplitude  and each producing phonons at the same  $2\bf q$ pseudomomentum (see figure~\ref{fig:sliding2}).    The simultaneous 2D production in emission or absorption with its 32 cm$^{-1}$ problem is thus alleviated (not to mention it is very weak compared to what sliding produces). The sliding mechanism also predicts the experimentally measured Stokes-anti-Stokes anomaly for 2D (see below).    The density of states for both the initial and final electronic states will have a major effect on the propensity to slide various amounts.    
 \vskip .07in \subsection{2D band  Stokes, anti-Stokes anomaly} \vskip .07in
 
The 2D Stokes, anti-Stokes anomaly is simple to explain.  In th sliding mechanism, both phonons share the same energy and diminishment, or augmentation (in anti-Stokes).  Each Stokes D (one in absorption, one in emission) suffers an 8.4 cm$^{-1}$ diminishment, or 17 cm$^{-1}$ total, and each anti-Stokes D enjoys an  8.4, cm$^{-1}$ augmentation, or 17 cm$^{-1}$ altogether, for an overall 34 cm$^{-1}$ asymmetry, four times greater, not the expected two times greater, than  the 8.4 cm$^{-1}$ D band asymmetry.  Again, this is  in agreement with experiment (see figure~\ref{fig:dressNoPerm}).
%
 
 The D and 2D Stokes, anti-Stokes anomalies  are thus easily explainable using KHD theory,  unlike the very elaborate rationale used in the  DR model\cite{dressNoPerm}. 
 
 \vskip .07in\subsection{D and 2D bandwidths; 2D intensity with laser frequency} \vskip .07in
 
The D-mode band must be  broader than the   band centers spaced by 8 or 8.4 cm$^{-1}$  that comprise it. We    find it to be  ca. 20-25 cm$^{-1}$,   FWHM in the literature, about twice the FWHM of G the mode, or 13 cm$^{-1}$\cite{Ferrari_ePh}.  The G  has no dispersion and no issues of Stokes,  anti-Stokes anomalies. A D  line much broader than 13 cm$^{-1}$ supports  the idea that the Stokes D mode is really the superposition of two displaced peaks, one coming from production in absorption, another in emission. Not only are the Stokes and anti-Stokes peaks  shifted by 8.4 cm$^{-1}$; they also must be overlapping when brought to the same side of the Rayleigh peak, exactly as seen in experiment; see also  figure~\ref{fig:dressNoPerm}.
 
 The 2D bandwidth is only somewhat larger than D, approximately 30 cm$^{-1}$.  It is not a double peak, at least not until the symmetry is broken (as revealed by tensioning the sample in some direction.)  It seems likely that 2D earns its width in a different manner than D,  perhaps a result of the sliding process on   slightly nonlinear or trigonally warped Dirac cones, or off-resonant sliding.

 The decrease in  2D intensity with increase laser energy, and broadening of 2D, leading to near complete absence at 266 nm in the ultraviolet\cite{liu2015deep},  is easy to explain based on the KHD and sliding transition picture.    The graphene electronic dispersion is suffering significant bending (as distinct from trigonal warping) as the ultraviolet is approached, degrading the linearity essential to the sliding 2D mechanism and its interference enhancement.  Note that is G mode is robust to the laser energy increase, and indeed it does not transition slide.  It has been reported to increase intensity at the fourth power if the laser frequency\cite{liu2015deep}, which incidentally fits the $\omega_I \omega_S^3$ classic KHD dependence, equation~\ref{KHD}.   G is ready to emit a Raman shifted photon immediately upon photoabsorption, and benefits from off-resonance contributions. The first D absorption step in 2D sliding needs to be close to on resonance, since the phonon created and the associates electron recoil cannot immediately emit, mitigating against a virtual process.

 \vskip .07in \section{Defect and Laser Frequency Trends} \vskip .07in
 
Several interesting trends develop  in the Raman spectra  as density of defects or laser frequency changes. 
We begin with the fascinating similarity of  sidebands in polyacetylene and graphene. High symmetry  $k=0$ dispersionless bands can be  parents of dispersive sidebands carrying momentum $2q_v^-$, coming from the production of a phonon in absorption, where $q^-$ is slightly less than $q$, the higher electronic  pseudomomentum  when no phonon is produced.  The conduction band electron gets kicked to $-q_v^-$ as it generates a $2q_v^-$ phonon.  It requires  elastic backscattering to  appear    in the Raman signal. In emission, the sideband phonons carry momentum $2q_v$. 

The trends in the  G$^\prime$ band   are quite parallel to sidebands  in polyacetylene: fixed $k=0$ peaks with  nearby dispersive sidebands, growing in intensity with increasing sources of elastic backscattering, and moving in frequency according to the band dispersion and phonon $q$.    The G$^\prime$ band (formerly called D$^\prime$) has nothing to do with the D band, and is simply the sideband to G, as Hilke \cite{hilke} and others have known for some time. 

\begin{figure}[htbp] 
   \centering
   \includegraphics[width=6.0in]{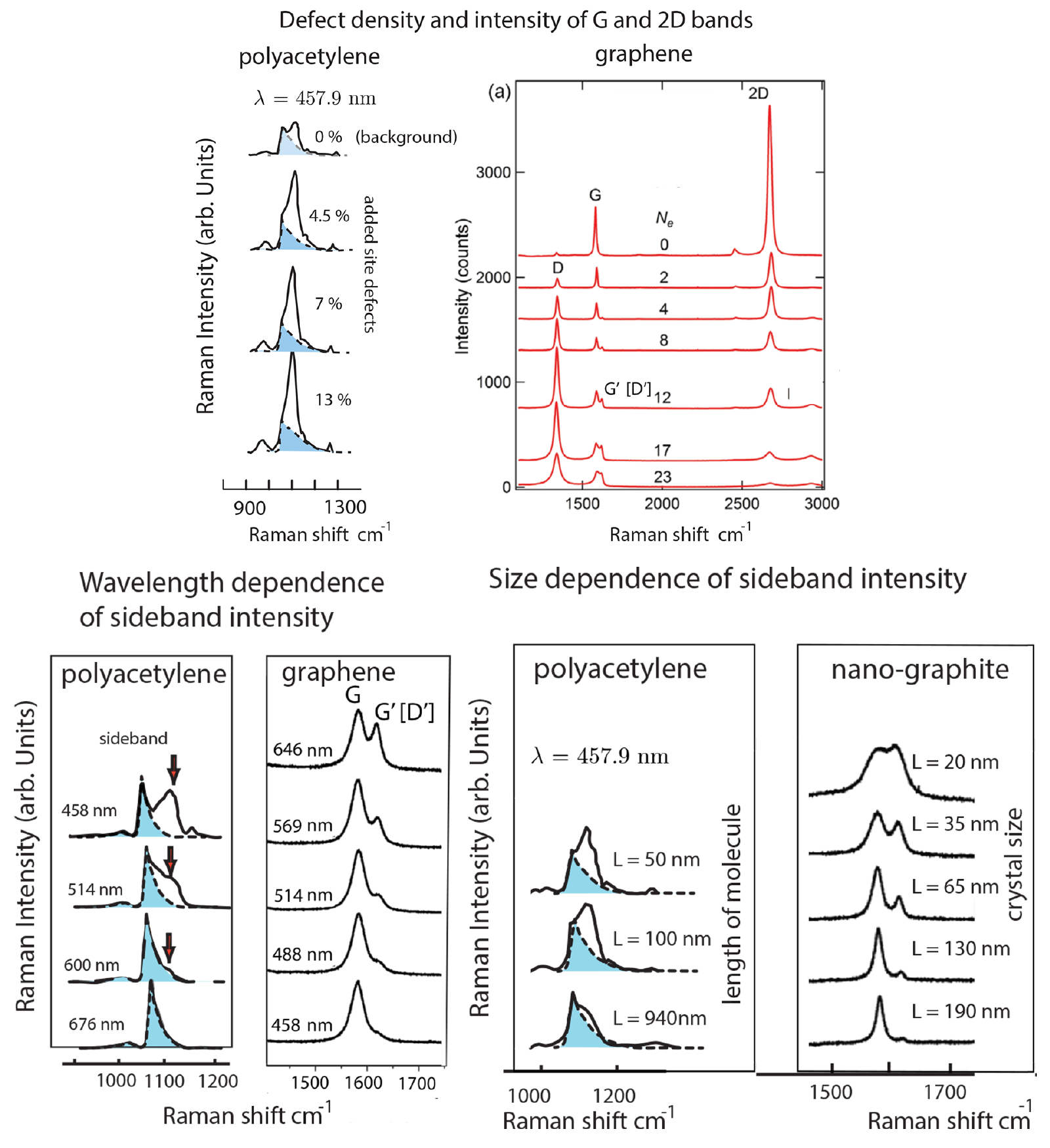} 
   \caption{(top) Response of polyacetylene and graphene Raman spectra to defects.  (top left)  The sideband growth is shown in the 1054 cm$^{-1}$ polyacetylene band as elastic backscattering defects are added. The figure was redrawn and the  $k=0$ band contribution colored, using Sch\"afer-Siebert  \textit{et. al.} \cite{defects}.  The sideband is present   at 0\% added defects because of pre-existing defects (including molecular ends).      The total intensity consists of the $k=0$ band and  the overlapping sideband, making a break in slope  to the right of the $k=0$ band  peak.   (top right) An important graphene Raman study taken from Childres   \textit{et.al.}\cite{childres2011effect}, with permission, with  similar sideband growth (G$^\prime$, sideband to G).  The   dramatic reduction in the 2D intensity with defect density is explained in the text. (bottom) Trends in sideband growth and dispersion are shown  for polyacetylene and graphene (red arrows in the case of polyacetylene, giving   band frequency predictions  based on electron and phonon dispersion \cite{heller2015raman}). The change with laser energy is opposite in the two cases.  }
   \label{fig:defectGraphene3}
\end{figure}

The growth in sideband intensity   with increasing sources of impurity backscattering is seen on the left of figure~\ref{fig:defectGraphene3} \cite{defects}. We explained these trends entirely within a KHD context applied to the one dimensional polyacetylene crystal with defects \cite{heller2015raman}.

We reproduce    studies of the  trends with laser energy in figure~\ref{fig:defectGraphene3}, left, for both graphene\cite{cancado2006general} and polyacetylene \cite{Kuzmany}.  In graphene, the conductance  trend is toward increased sideband strength with decreasing laser energy, as seen in figure~\ref{fig:defectGraphene3}. In polyacetylene, the trend is reversed.  One obvious difference is that a propagating electron wave cannot fail to collide with  any defect in one-dimensional polyacetylene. We have undertaken density functional calculations on polyacetylene distorted by kinks and other geometrical defects, which  show that higher energy electronic states backscatter more readily from such defects than do lower energy ones,  in the conduction band.  
   \vskip .07in
\subsection{The D and 2D bands are sidebands to ``D$^K$'',   a forbidden k=0 K point vibration}
\vskip .07in
  In  a definable sense, the D and 2D bands are   dispersive sidebands to a forbidden   k=0 Dirac K point phonon we call D$^K$, living at the vibrational K-point.  In analogous cases, such as the G$^\prime$ [D$^\prime$] and 2G$^\prime$ [2D$^\prime$] bands  of graphene (dispersive sidebands to the non-dispersive G band), see figure~\ref{fig:defectGraphene3},  the dispersive sidebands in polyacetylene \cite{heller2015raman}  are partner to visible non-dispersive bands.  In those cases the non-dispersive bands belong to $\Gamma$ point vibrations.  The band edge   $K$ point  vibration  D$^K$ is   of course not a   $\Gamma$ point vibration; one consequence is that it has a vanishing transition moment and Raman intensity, as is required by symmetry.  This however  does not disqualify the D and 2D bands from being labeled dispersive sidebands to this silent and  dispersion-less  parent band;  it would  lie  typically 8 cm$^{-1}$ to the left of D and not require backscattering.   

 \vskip .07in\subsection{Defects and 2D intensity } \vskip .07in
There is  a well known dramatic  \textit{decrease} in the 2D band intensity with increased defect density \cite{Cooper} (figure~\ref{fig:defectGraphene3}).  This is easily understood in terms of the sliding production of the two D phonons. The amplitude for production  of the first D phonon in absorption is relatively insensitive to defect density.  Transitions with no sliding  contribute to  D Raman intensity if they are elastically backscattered, and indeed the D intensity increases with added impurities. Sliding D transitions  are Pauli blocked, even if backscattered. Sliding transitioned electrons are equally prone to defect elastic backscattering or more general scattering,  making it extremely unlikely they can produce another D in emission, being unable to reverse   absorption path.  Defect scattering of the conduction electron  thus quenches the source of  D phonon production in emission, and   the 2D Raman band intensity diminishes with it  as defects are added.  This is just what is seen in figure~\ref{fig:defectGraphene3}.    
 The reverse trends of D and 2D intensity with added defects therefore follow from the sliding, ``up and back down along a reversed path" mechanism for  2D Raman emission.
\vskip .07in
\subsection{Anomalous spacing of D and 2D} 
\vskip .07in
   A   prediction can be made about the D, 2D spacing seen in experiments at any frequency. This has been discussed within the DR model also \cite{dressNoPerm,ding2010stretchable}. The frequency of 2D is smaller than twice D, by amounts depending on experimental conditions.  The ideal ``bare,'' unstrained,  low temperature, and fairly clean (but dirty enough to see D) graphene experiment has not been done to our knowledge.  However, quoting Ding   \textit{et.al.} \cite{ding2010stretchable}, ``The results show that the D peak is composed of two peaks, unambiguously revealing that the 2D peak frequency ($\omega_{2D}$ ) is not exactly twice that of the D peak ($\omega_D$ ). This finding is confirmed by varying the biaxial strain of the graphene, from which we observe that the shift of $\omega_{2D/2}$  and $\omega_D$  are different.''
   
   According to our application of KHD to graphene, a 1335  cm$^{-1}$  D phonon   produced in absorption is diminished owing to the 50  cm$^{-1}$/eV phonon dispersion, by 1335/8065 X 50 = 8.28 cm$^{-1}$.  A  D phonon produced in emission is undiminished.    The two bands will overlap to make a broader feature than either component.  Assuming the two bands are equally intense, as KHD predicts, there should be a  combined band with an average 4 cm$^{-1}$ displacement to the left in the Stokes spectrum.   The idea that  D is composed of two bands with an 8 cm$^{-1}$ splitting was also suggested within the  DR model \cite{dressNoPerm}, with a very   different  justification, ``depending on which of the intermediate states is virtual'' \cite{ferrariNat}. 
The reason for the two bands is actually much less exotic (absorption \textit {vs.} emission production) and on a firmer foundation in KHD (both real, resonant processes)  than the virtual processes required in DR.

    The 2D band consists of two separately produced, diminished D phonons.  There is an  8 cm$^{-1}$ diminishment in absorption, and  a matched   8 cm$^{-1}$ diminishment in emission according to the sliding scenario, totaling a 16 cm$^{-1}$ shift.  As just discussed the D band is displaced by 4 cm$^{-1}$, thus twice the frequency of D is predicted to be 8  cm$^{-1}$ shifted as opposed to the  16 cm$^{-1}$ shift of 2D, or a  -8 cm$^{-1}$ difference between 2$ \times $D and 2D.  Review of many published spectra under different conditions shows  $E(2D) - 2 E(D) \sim   - 2  \ \ \textrm{to } -10  \textrm{ cm}^{-1}$.  However   samples were suspended on different substrates by a variety of methods; measured shifts depend on these conditions. Reference \cite{ding2010stretchable} shows  that any source of stretching or compression can affect the D, 2D distance.    The D, 2D  shift deserves more investigation using suspended, gently pinned graphene.
 

 \vskip .07in \section{Mixed Bands and Bandwidth Trends } \vskip .07in
 \begin{figure}[t] 
  \centering
   \includegraphics[width=4.in]{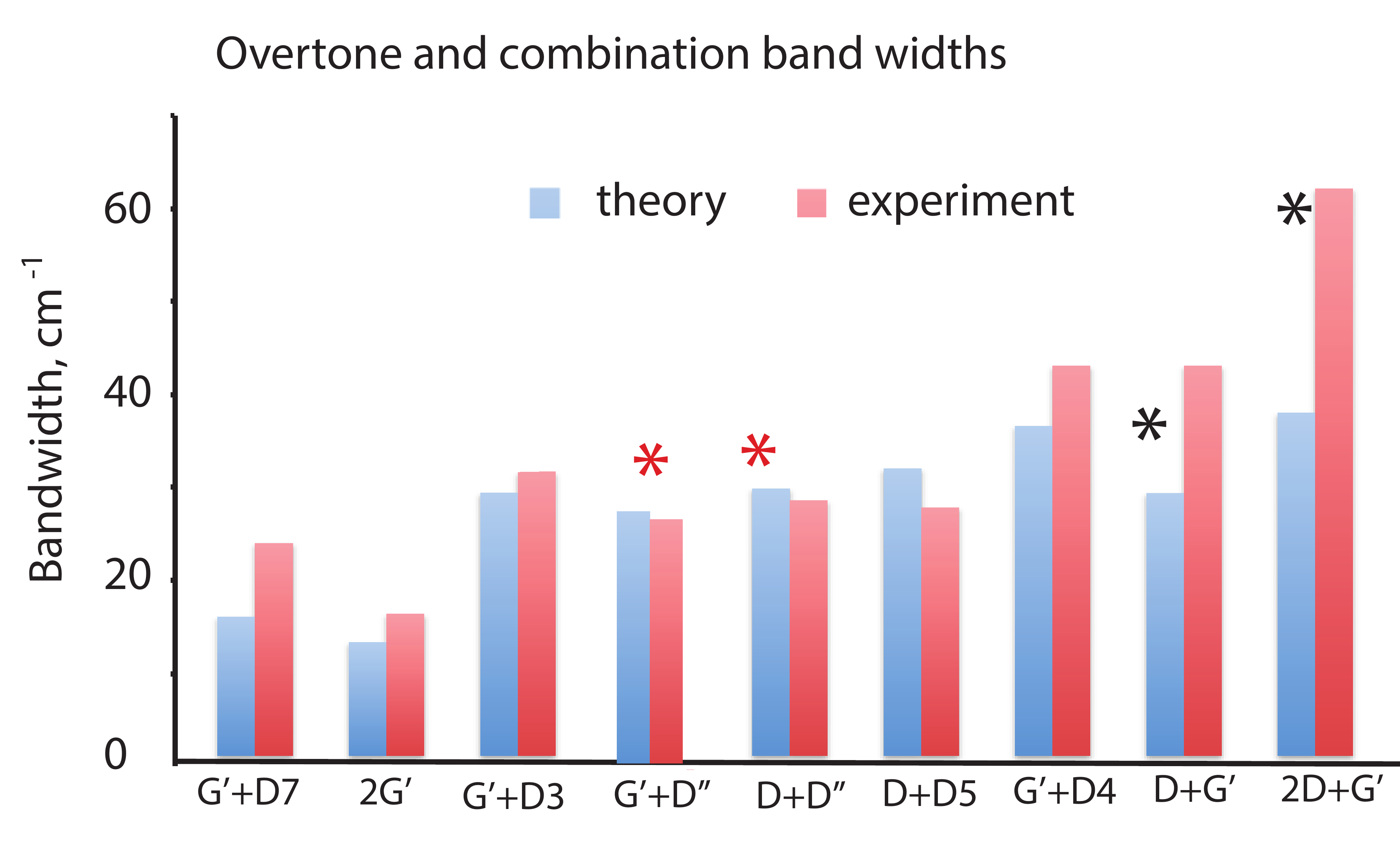} 
   \caption{Calculated (blue) and experimental (from \cite{hilke}, red) bandwidths for mostly mixed combination bands. The calculations were as follows: The wavenumber difference for G$^\prime$  (old D$^\prime$ ) up first and D4 down  \textit{versus}   D4  up first and G$^\prime$ down was calculated using the mode frequencies and dispersion slopes from  \cite{hilke}. 
      The Raman shift (and phonon energies) depend on which transition   is  first, and this difference is widening the bands.    For modes $a$ and $b$, the formula is $W = (\vert S_a-S_b\vert ) (\hbar \omega_a + \hbar \omega_b)$ where $W$ is the  component of the width coming from which mode is created first,  $S_i$ is the dispersion slope in cm$^{-1}$ per eV, and $\hbar \omega_i$ is the energy of the phonon in eV.   
 The theoretical results are compared to widths found in the data from  \cite{hilke} after adding 16 cm$^{-1}$ for the ``intrinsic'' width.   Except for the G$^\prime$ + 2D  (assignment and data taken from\cite{rao2011multiphonon}) and G$^\prime$ + D, the agreement is  good, and it lends confidence to  the KHD approach and the one up, one down scenario double phonon. See text for discussion of bands labeled by a red star.}
   \label{fig:width}
\end{figure}
Does the   sliding   mechanism enhancing the 2D band brightness also contribute to the strength of other bands?  We have already noted that  sliding does not help the D band  gain intensity, since the electron is Pauli blocked even with elastic backscattering (however, see the last section  for the changed situation when the sample is hole doped). The G band transition  cannot  slide   since the $\Gamma$ point  produces  a $k=0$ phonon. The same applies to 2G, which does not appear in the Raman spectrum regardless of defects.  

The sliding mechanism for mixed transitions is a different story than for homogeneous ones, especially for bandwidth. The momentum conservation requirements on the production of a  pair of phonons requires  that they are matched in $\bf q$.  They do not need to be matched in  type; e.g. a Raman band for G$^\prime$ (old D$^\prime$) and  D$^3$ (Hilke's notation) could be produced by the sliding mechanism.   This fact  allows us to explain many of the disparate bandwidths seen in the Hilke  \textit{et.al.} data \cite{hilke}, since  differences in dispersion and frequency of the two components contributes to the bandwidths, as we presently see. 

Most of the  mixed transitions   are  weak (and certainly would be invisible without transition sliding), as figure~\ref{fig:tracehilke} shows. The weakness may reflect  small transition matrix elements or the possibility of destructive interference between terms in the sum in  equation \ref{KHD}.  
The strength of  homogeneous two phonon transitions, like 2D or 2G$^\prime$, is expected to be high since they are produced by mirror image processes with in-phase numerators.  

  For mixed transitions, sliding applies but leads to slightly different Raman band frequencies depending on which phonon is produced first, in absorption.  The bandwidth will reflect this (see figure~\ref{fig:width}).  Using data  from Hilke \cite{hilke} at 288 nm, we arrive at figure~\ref{fig:width}. The reasons for the bandwidths of the D+  G$^\prime$ and 2D +  G$^\prime$ are discussed below.  
Red stars on D + D$^{\prime\prime}$ and G$^\prime$ + D$^{\prime\prime}$: We have used reference~\cite{may2013signature}  to help understand the skewed line at about 2450 cm$^{-1}$ with the nominal assignment D + D$^{\prime\prime}$.    This study decomposed it into two bands, one of which is D + D$^{\prime\prime}$ with a width of 20 cm$^{-1}$, and the overlapping  higher energy LO G$^\prime$ band   near the $\Gamma$ point, but now near the K point, a   less intense band, with a FWHM of 29 cm$^{-1}$.  

Some mixed sliding transitions, such as G$^\prime$ + D$^3$, G$^\prime$ + D$^4$, and even some hint of D + D$^5$ (Hilke's notation, except D$^\prime \to$ G$^\prime$) do not require defect backscattering and are seen weakly in high quality spectra of clean graphene, for example in Childres \textit {et. al.}, reference \cite{childres2013raman}.  

  Consider a mixed overtone band involving phonon modes A and B.
If A is created in absorption,   the  $\bf q_A^-$ of the transition will be different than if B is created first, giving an electron pseudomomentum $\bf q_B^-$.  If A is a higher frequency phonon than B, the electronic transition energy diminishment is larger if A appears in absorption,  and   $\bf q_A^-$ will be smaller.  The emission B phonon must follow with the opposite  pseudomomentum $2\bf q_A^-$.  This allows for matched electron-hole recombination, but the B phonon is required to adjust its energy to arrive at the  right $\bf q_A^-$. This energy correction depends on its momentum dispersion.  Thus the total phonon energy (and thus the Raman shift) is slightly different if A is created in absorption than if B is. This fact contributes to a small uncertainly in frequency and contributes to the band width. We calculated the  bandwidths for each mixed transition and added 16 cm$^{-1}$  to allow for the intrinsic broadening seen  in   narrow  bands, due presumably to  phonon decay \cite{lazzeri2006phonon,pisana2007breakdown}. 
Using data mostly from Hilke \cite{hilke} at 288 nm, we arrive at figure~\ref{fig:width}. This gave rise to figure~\ref{fig:width}.

\vskip .07in
\noindent \section{Evidence of Sliding D Absorption}
 \vskip .07in 
 \label{evidence}
   Chen and co-workers\cite{chenCrommie} hole doped the valence band by as much as 0.8  eV and saw abundant continuum emission,  in a certain range of depletion and Raman shift. We now show this emission, that as reference \cite{chenCrommie}  points out, integrates to  more than 100 times the strength of the G band that it overlaps, can be explained by sliding D phonon transitions that are normally Pauli blocked  or could have been the first step in making a 2D pair by reversing the sliding transition, coupled with an electronic Raman component.  
  
\begin{figure}[htbp] 
   \centering
   \includegraphics[width=6.6in]{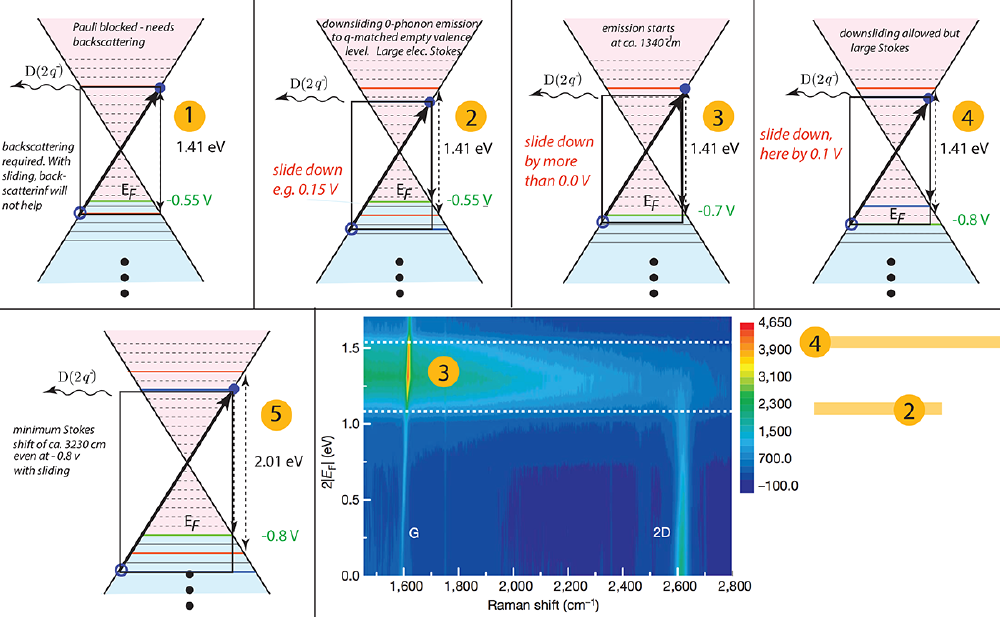} 
   \caption{Continuum emission in graphene from   normally Pauli blocked sliding   transitions at some laser energies and hole dopings. At the top, four scenarios for doping levels and sliding transitions at 1.58 eV laser energy are shown, together with their relation to the data of  reference~\cite{chenCrommie}. For a D phonon, the diminished electronic part of the transition is at 1.41 eV = 1.58 eV (laser energy) - 0.166 eV (phonon energy). Scenario 1 at $2 \vert E_F\vert = 1.1$ eV is Pauli blocked without defect backscattering and therefore invisible in Raman.  Scenario 2 at $2 \vert E_F\vert = 1.1$ eV  is shown sliding down 0.15 eV, thus matching the hole and avoiding Pauli blocking.  However is  produces a minimum 3765 cm$^{-1}$ Stokes shift (lower dashed line). Scenario 3 at $2 \vert E_F\vert = 1.4$ eV is within the bright continuum emission starting at ca. 1340 cm$^{-1}$ and going higher, but cut off at  higher Stokes shifts by (1) declining density of states as the relevant conduction and valence bands  approach the Dirac point,   and (2) the emission factor $\omega_s^3$ (see equation~\ref{KHD}).  Scenario 4 (upper dashed line) needs to slide down to reach occupied levels that can be promoted to the conduction band, but this again causes large Stokes shifts. Scenario 5 applies to a hypothetical  2.18 eV laser energy (not used in the hole doping experiment of reference~\cite{chenCrommie}); it shows that at any available hole doping, the D + electronic Raman shift continuum emission would not appear anywhere near the low energy regime seen in the current experiment.   The complete fading out of the 2D band at the onset of the continuum emission (lower right)  as $2\vert E_F\vert$ is increased is explained in the text. The panel at lower right is re-drawn from reference~\cite{chenCrommie}. }
   \label{fig:blob}
\end{figure}

   With hole doping,      D phonons produced  in a sliding transition, and the associated electronic conduction band level are normally orphaned by Pauli blocking (unless the electron returns by the sliding 2D mechanism along the reverse path).    Now the electron has  a new option: to emit  to a    valence orbital emptied by doping \textit {without creating another phonon}. The original hole remains unfilled,  and the valence orbital      is filled instead.    This    leads to a continuum of  potentially large  electronic Stokes shifts. (Electronic Stokes because the valence electron has  been promoted, not  phonon in the final step.) This  phonon-less emission channel  is far more likely than creating another phonon, and thus the ``feedstock'' of the 2D band is depleted, quenching the 2D band.  Thus the 2D band   should fade out in the experiment  as the continuum  emission appears, just as seen in the experiment (see lower dotted white line,  figure~\ref{fig:blob}). Before that happens, we note that the deeper the hole doping, the less sliding up distance is available before the last populated initial valence level is reached. Deeper doping progressively removes sliding 2D transitions, weakening the 2D band as $2\vert E_F\vert$ increases.  The consistent, progressive weakening of the 2D starting as low as  $2\vert E_F\vert=0$ is evident in the lower right experimental data.

 \vskip .07in
\noindent \subsection{G mode brightening}
\begin{figure}[htbp] 
    \centering
    \includegraphics[width=6.6in]{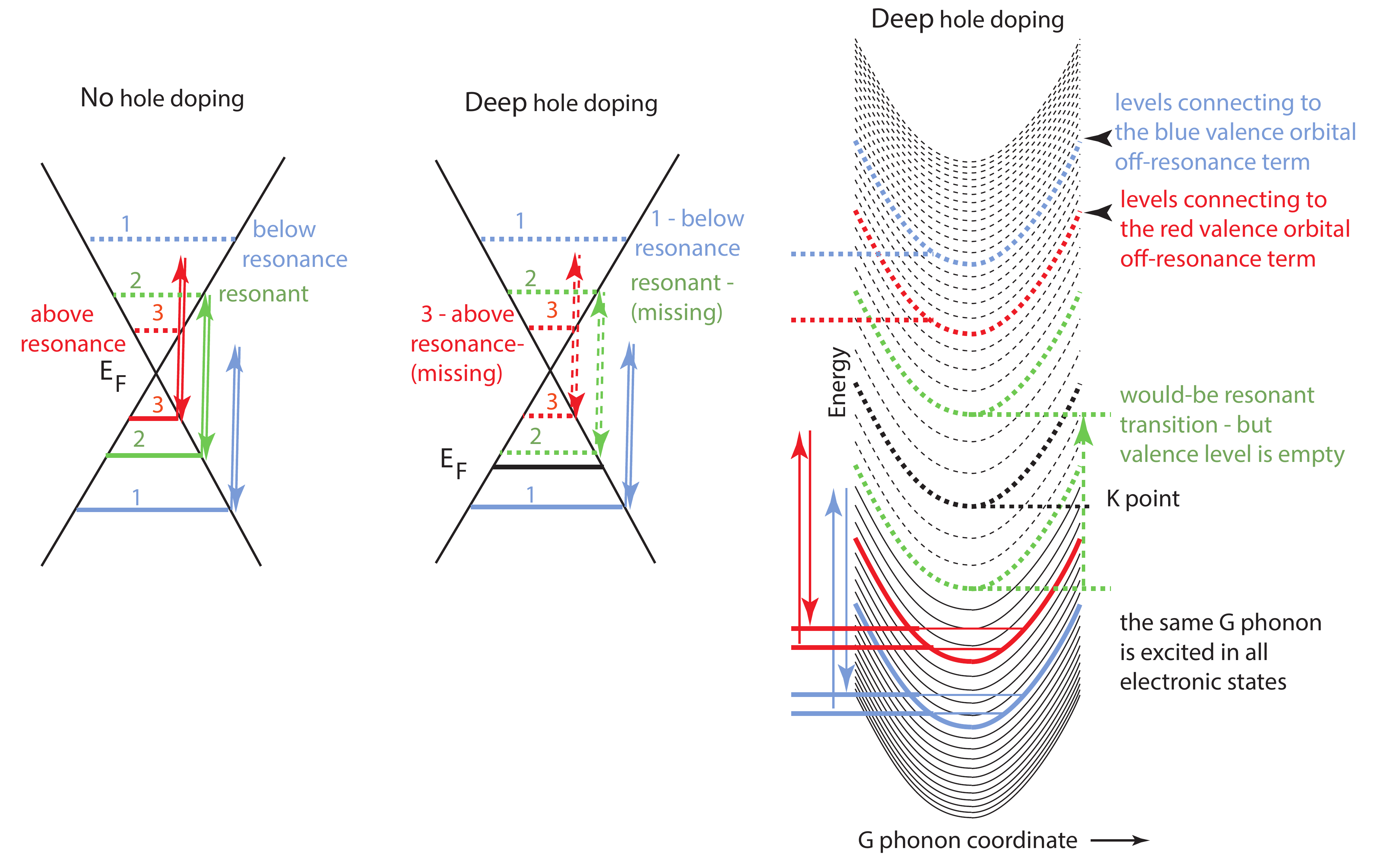} 
    \caption{(left) Non-resonant (and one resonant)  electronic  processes that can contribute to G mode creation. In the case labeled 1, the below resonance  electronic transition (of the type seen in  off-resonance Raman scattering) transiently promotes the electron from the lower blue solid to the upper  blue dashed line.  There is nothing available  (with a G phonon produced) well above or below the upper blue dashed state (which is an electron-hole-phonon triplet) with a nonvanishing matrix element. Process 2 is on-resonance production of a G phonon. Process 3 is above resonance with   red solid line representing the initial valence state and red dashed line  indicating the state to which a  virtual transition has a nonvanishing transition moment.    All three cases apply to the same incident photon.   (middle) With hole doping, some  transitions are no longer possible, if the initial valence state is empty.  (right)  The G mode phonon coordinate is displayed  as the abscissa,   energy  the ordinate.  Filled bands are solid; empty bands are dashed.   The density of states is indicated qualitatively using the level spacing. The restriction red-to-red, blue-to-blue is due to q-conservation (small energy differences due to phonon production not shown).  A continuum of off-resonant transitions apply for the same laser energy (arrows are the same lengths in red, blue, and   excluded green examples) and the same phonon.  The sum over such states is truncated here compared to the no doping case, but the missing terms  have  the opposite sign in their real part, above resonance.  Thus hole doping enhances G, for reason of removal of destructive interference,  in agreement with reference~\cite{chenCrommie}. Time-energy arguments apply: the time spent in the excited state is $\hbar$ times the inverse of its energy deficit from resonance.   }
    \label{fig:nonres}
     \end{figure}

 \vskip .07in 
 \label{bright}
 In the paper by Wang and his group\cite{chenCrommie}, a brightening of the G band is noted as hole doping is increased.  It can be seen as a gradual waxing of G  intensity in figure~\ref{fig:blob}, even after the continuum band is exceeded in the upper left corner of the plot on the lower right. The authors attributed this to removal by doping of destructively interfering paths.  This also happens within the KHD picture.  The sum over nonresonant states $\vert n\rangle$ of energy $E_n$ normally extends above and below resonance, which causes cancellations in the real part of the sum.  Doping eliminates part of the sum on the high side of resonance, enhancing the real part.  The relevant states $\vert n\rangle$ all contain matched electron-hole G phonon triplets and matched electron-hole pairs (if G is to be produced in emission, with slightly different energy denominator) even if $E_n$ is quite non-resonant (see figure~\ref{fig:nonres})
 
    When present, real (not virtual) processes play a dominant role in KHD, and apart from hole or particle doping scenarios real pathways are always available in   graphene.  Virtual processes  (such as those present in ordinary off-resonant Raman scattering)   do not normally play a center stage role, living  in the shadows of the real, resonant processes,  contributing mostly near resonance, in accordance with damping factors.  
    
 Of course, off-resonant (pre- or post-resonant) Raman scattering still operates in graphene as it does in molecular systems. For hole doped graphene such transitions may dominate when the resonant initial valence states are depleted of population. For example, starting on the lower Dirac cone below a hole doped, lowered $E_F$, a laser may be too low in energy  to reach resonant levels on the upper cone (see figure~\ref{fig:nonres},  middle); yet,  electron-hole excitation and recombination with no Pauli blocking quickly follow upon virtual absorption, on a timescale $\Delta t$ given by the detuning $\Delta E$ from resonance, where $\Delta t\Delta E \sim \hbar/2$. A phonon may thus be created or destroyed, off resonance.  Again, the coordinate dependence of the transition moment is responsible.  Time can become too short for electron-phonon scattering or any nuclear motion  to develop, even though Raman scattering is quite robust. Raman intensity must come instead from an instantaneous phonon creation/annihilation process, which KHD provides through the transition moment coordinate dependence.

\section{Conclusion}
The  universally  used (except in the  conjugated carbon community), 90 year old KHD Raman scattering theory has been applied for the  first time in graphene Raman scattering theory. The results are in excellent agreement with experiment, and new insights (and predictions about new hole doping experiments) have resulted.  The most important of these may be the sliding transitions, explaining the brightness of overtone transitions in graphene. 

If KHD is used  with constant transition moments (without justification),  the  resulting lack of any predicted Raman scattering in graphene  might induce one   look elsewhere.  ``Elsewhere'' ought to have been to free up the transition moments,  but DR instead keeps them constant, and furthermore bypasses KHD completely. Instead,  \textit{ad hoc},  one or two more orders of perturbatively  treated  electron-phonon scattering are inserted to finally create  phonons in the conduction band. The result is a  fourth order perturbation theory, two orders beyond KHD.  It is hard to see why these phonons should not be 100\% Raman inactive due to Pauli blocking of the radically changed conduction band electronic levels after inelastic scattering.  Nontheless  DR   leads to a parallel universe so to speak,  if one is willing to turn a blind eye to Pauli blocking, that can  be used as a model to catalog observed phenomena.  The underlying physics is entirely different than KHD, in fact it is physically incorrect. Since the  proven, traditional, lower order (in perturbation theory)  KHD mechanism works extremely well, the DR narrative must in retrospect be regarded as an outlier and a model at best.

\vskip .2in
 \textbf{Acknowledgements} The authors acknowledge support from the NSF Center for Integrated Quantum Materials(CIQM) through grant NSF-DMR-1231319. We thank the Faculty of Arts and Sciences and the Department of Chemistry and Chemical Biology at Harvard University for generous partial support of this work. We are indebted to  Profs. Michael Hilke, Feng Wang,  Yong  Chen, and Philip Kim for helpful discussions and suggestions. We also thank Fikriye Idil Kaya for suggestions and a careful reading of the manuscript. We are grateful to V. Damljanovic for correcting an error concerning the multiplicity of the $\Gamma$ point G mode.
 
 \bibliographystyle{unsrt}
  \bibliography{Raman}

\section{Supplementary Material}

              \vskip .07in
\subsection{ {Time domain KHD}}
\vskip .07in
 \label{TDPT}
According to the  time dependent form of   KHD theory \cite{leeheller1,heller2}, the amplitude to scatter from phonon state ${\bf m}_i$ is 
\begin{eqnarray}
\label{make2}
a_{{\bf m}_i{\bf m}_f}^{\rho,\sigma}(t)  
&=&\!
    \int\limits_{-\infty}^t \! dt'\!\!\int\limits_{-\infty}^{t'} dt''\ 
\langle \psi_{{\bf m}_f}^{B.O.} \vert  {\bf G}_{0}^+(t-t') \ {\bf V}(t')\ 
{\bf G}_{0}^+(t'-t''){\bf V}(t'') {\bf G}_{0}^+(t'')\vert \psi_{{\bf m}_i}^{B.O.} \rangle   \\
&=&\!
  \int\limits_{-\infty}^t \! \! dt'\!\!\!\int\limits_{-\infty}^{t'}\! \!dt''\ 
\langle \psi_{{\bf m}_f}^{B.O.} \vert  e^{iE_{{\bf m}_f}(t-t')/\hbar} \ \    {\bf D}^\sigma\  \ 
e^{-i{\bf H}^{B.O.}(t'-t'')/\hbar -\Gamma(t'-t'')/\hbar} \beta(t'')\ {\bf D}^\rho \   e^{-iE_{{\bf m}_i} t''/\hbar}\vert 
\psi_{{\bf m}_i}^{B.O.} \rangle  \nonumber
\end{eqnarray}
where 
${\bf D}^\rho$ is the dipole operator  polarization $\rho$.
  We have incorporated a damping factor $\Gamma$ to account for the environmental factors not explicitly included in the Hamiltonian. ${\bf G}_{0}^+(t'-t'') = e^{-i{\bf H}^{B.O.}(t'-t'')/\hbar}, t' >t''; \  {\bf G}_{0}^+(t'-t'') = 0, t' <t''$  is a retarded Born-Oppenheimer Green function (one that propagates Born-Oppenheimer eigenstates  unchanged except for a phase factor) and ${\bf V}(t')$ is the light-matter perturbation with arbitrary time dependence governed by $\beta(t'')$, which we take normally to be $\exp[i\omega_I t'']$ corresponding to a cw laser.

The expression~\ref{make2} shows clearly that the propagation on the conduction band Born-Oppenheimer potential surface takes place after the transition moment ${\bf D}^\rho$ has acted at time $t''$ on the initial valence wave function.  The transition moment changes the functional form of that wave function and the Born-Oppenheimer Hamiltonian is presented at time $t''$ with newly created  or destroyed phonons relative to the valence state, before any excited state propagation has taken place.  This is also clear below  after we insert a complete set of Born-Oppenheimer eigenstates to resolve the propagator. The excited state propagator acts until time $t'$, when the electron fills the hole, giving the  transition moment  another chance to act.  \textit {No phonons are created or destroyed during the time evolution  in the conduction band, according to KHD, nor are they needed to produce the Raman signal.}

$\vert\psi_{{\bf m}_i}^{B.O.}({\boldsymbol \xi};{\bf r}) \rangle$ $  = \vert\phi({\boldsymbol \xi};{\bf r})\rangle\vert \chi_{{\bf m}_i}({\boldsymbol \xi})\rangle$ is the Born-Oppenheimer state before the photon interacts.  The electronic state $\phi({\boldsymbol \xi};{\bf r})$ is a function of all electrons at {\bf r}, depending parametrically on all the phonon coordinates $\boldsymbol \xi$ . The dipole moment connecting the initial electronic state $i$ and the electron-hole pair state $\bf{q}_c,\bf{q}_v$,  $\mu_{\bf{q}_c,\bf{q}_v}^i({\boldsymbol \xi})$, is a function of the phonon coordinates  ${\boldsymbol \xi}$, defined as
\begin{equation}
\mu_{ \bf{q}_c,\bf{q}_v}^\sigma({\boldsymbol \xi}) =   \langle\phi_{\bf{q}_c}({\boldsymbol \xi}; {\bf r})\vert {\bf D}^\sigma \vert\phi_{\bf{q}_v}({\boldsymbol \xi};{\bf r})\rangle_{{\bf r}},
\end{equation}
 where the subscript $\boldsymbol r$ reminds us that only the electronic coordinates are integrated. We insert  a complete set of  Born-Oppenheimer  eigenstates (they are complete, even if not exact eigenstates of the full Hamiltonian) in front of the Born-Oppenheimer  propagator in equation \ref{make2}:
\begin{equation}
\label{insert}
1 = \sum\limits_{c,v,\bf m} \vert\phi_{\bf{q}_c\bf{q}_v}({\boldsymbol \xi};{\bf r})\rangle\vert \chi_{\bf m}({\boldsymbol \xi})\rangle\langle\chi_{\bf m}({\boldsymbol \xi})\vert \langle\phi_{\bf{q}_c,\bf{q}_v}({\boldsymbol \xi};{\bf r})\vert.
\end{equation}
where we have acknowledged  by absence of a subscripts on the phonon wave function  $\chi_{\bf m}({\boldsymbol \xi})$ that the phonon modes do not change upon electron-hole pair formation (in extended systems like graphene).
Since $H^{B.O.}\vert\phi_{\bf{q}_c,\bf{q}_v}({\boldsymbol \xi};{\bf r})\rangle\vert \chi_{\bf m}({\boldsymbol \xi})\rangle = E_{c,v,{\bf m}} \vert\phi_{\bf{q}_c,\bf{q}_v}({\boldsymbol \xi};{\bf r})\rangle\vert \chi_{\bf m}({\boldsymbol \xi})\rangle$,    we have 
\begin{eqnarray}
\label{make3}
a_{{\bf m}_f{\bf m}_i}^{\rho,\sigma}(t)  
&=&\!
 \sum\limits_{c,v,\bf m_{vc}}  \int\limits_{-\infty}^t \! \! dt'\!\!\!\int\limits_{-\infty}^{t'}\! \!dt''\ 
 e^{iE_{v,c,{\bf m}_f}(t-t')/\hbar}  
e^{-iE_{v,c,{\bf m}} (t'-t'')/\hbar -\Gamma(t'-t'')/\hbar}e^{i\omega_I t''} \ e^{-iE_{{\bf m}_i} t''/\hbar}\nonumber
\\
&\times & \langle \chi_{{\bf m}_f}({\boldsymbol \xi})  \vert \mu_{\bf{q}_i, \bf{q}_j  }^\sigma({\boldsymbol \xi}) 
 \vert \chi_{{\bf m}}({\boldsymbol \xi})   \rangle \langle  \chi_{{\bf m}}({\boldsymbol \xi)}\vert \mu_{\bf{q}_c,\bf{q}_v  }^\rho({\boldsymbol \xi}) \   
\vert \chi_{{\bf m}_{vc}}({\boldsymbol \xi})   \rangle.  
\end{eqnarray}

Apart from pre-factors, equation~\ref{make2} with the  insertion  equation~\ref{insert} can be easily converted (ignoring the second, off-resonant term as usual and gently damping  the laser field at infinite positive and negative times) to the Raman scattering amplitude for the ${\bf m}_i\to {\bf m}_f$ process starting and finishing on electronic state $i$ with incoming light of frequency $\omega_I$, incoming polarization $\rho,$ and outgoing $\sigma$, i.e. equation~\ref{KHD2}.

The final state is designated, apart from initial and final polarization, by the initial (and final) ground,  valence electronic state labeled by $i$ and the final phonon occupations labeled by ${\bf m}_f$.  The sum labeled by $c,v$ and ${\bf m}$ is over all electron-hole states and phonon occupations that connect both initial and final states \textit{via} the transition dipole ${\bf D}$. Here reside some surprising  and important terms, including the sliding transitions (see main text).

It is illustrative to  incorporate the transition moment into  the phonon wave function $\varphi_{\bf{q}_c  ,\bf{q}_v;{\bf m}}^\rho({\boldsymbol \xi})  =
 \mu_{\bf{q}_c,\bf{q}_v}^\rho({\boldsymbol \xi})\chi_{{\bf m}}({\boldsymbol \xi}) $.
Phonon excitations are included in $\varphi_{\bf{q}_c,  \bf{q}_v;{\bf m}}^\rho({\boldsymbol \xi})$ (the ``electron-hole-phonon triplets'') but may be much less common than the pure electron-hole pair amplitude. 

Equation~\ref{KHD2} can be returned usefully to a new time domain expression \cite{leeheller1},
\begin{eqnarray}
\label{time}
\alpha_{{\bf m}_f{\bf m}_c}^{\rho,\sigma}(\omega_I) &=& \frac{i}{\hbar}\sum\limits_{c,v}\int\limits_0^\infty e^{i(\omega_I + E_{i,{\bf m}_f})t/\hbar - \Gamma t/\hbar} \langle \varphi_{ \bf{q}_c, \bf{q}_v ;{\bf m}_f}^\sigma({\boldsymbol \xi}) \vert e^{-i {\bf H}^{B.O.}t/\hbar}\vert\varphi_{\bf{q}_c , \bf{q}_v; {\bf m}_i}^\rho({\boldsymbol \xi})\rangle  \ dt \nonumber \\
&\equiv&   \frac{i}{\hbar}\sum\limits_{c,v}\int\limits_0^\infty e^{i(\omega_I + E_{i,{\bf m}_f)}t/\hbar - \Gamma t/\hbar} \langle \varphi_{ \bf{q}_c ,\bf{q}_v ;{\bf m}_f}^\sigma({\boldsymbol \xi}) \vert \varphi_{\bf{q}_c , \bf{q}_v; {\bf m}_i}^\rho({t,\boldsymbol \xi})\rangle \ dt
\end{eqnarray}
It is important to note that the KHD Raman amplitude  is overall 2$^{nd}$ order, involving only  perturbation  in  the matter-radiation interaction. One could go to higher order by adding well known non-adiabatic correction terms to Born-Oppenheimer theory, but we do not do that here.  Even better, degenerate perturbation theory involving the same correction terms might be used to account for Kohn anomalies and possibly other effects. 

The sum labeled by $j$  is over all electron-hole states that connect both initial and final states \textit{via} the two transition dipoles ${\bf D}$. Here reside some surprising  and important terms, including the sliding transitions described in the main text.  Equation~\ref{time} is useful for many things, including  understanding the  effect of $\omega_I$ on the Raman amplitude, if something is understood about the time dependence of  the amplitude $\langle \varphi_{ \bf{q}_i, \bf{q}_j ;{\bf m}}^\sigma({\boldsymbol \xi}) \vert \varphi_{\bf{q}_j  ,\bf{q}_i; {\bf m}}^\rho({t,\boldsymbol \xi})\rangle $, especially for early times. The faster the amplitude grows in time, the more robust it will be  against  $\omega_I$  lying far from resonance for a given electron-hole state $p$.  This is due to the half Fourier transform aspect of the time integral and transient behavior near  $t=0$. Transition moment coordinate dependence permits, at time $t=0$,  
$\langle \varphi_{ \bf{q}_i, \bf{q}_j ;{\bf m}}^\sigma({\boldsymbol \xi}) \vert \varphi_{\bf{q}_j , \bf{q}_i; {\bf m}}^\rho({0,\boldsymbol \xi})\rangle\ne 0 $, i.e. immediate finite amplitude at $t=0$.  

Returning to the energy domain, we probe the effect of setting the transition moments constant in phonon coordinates:


\begin{eqnarray}
\alpha_{{\bf m}_f{\bf m}_i}^{\rho,\sigma}(\omega_I) \nonumber &=& \sum\limits_{c,v,{\bf m}} \  \frac{\mu_{\bf{q}_c, \bf{q}_v}^\rho \ \mu_{\bf{q}_c,\bf{q}_v}^\rho\langle\chi_{{\bf m}_f}({\boldsymbol \xi})  \vert \chi_{{\bf m}}({\boldsymbol \xi})\rangle\langle\chi_{{\bf m}}({\boldsymbol \xi})\vert\chi_{{\bf m}_i}({\boldsymbol \xi})\rangle}{(\hbar\omega_I + E_{i,{\bf m}} -E_{c,v,{\bf m}}  + i\Gamma )}\\
\nonumber &=& 0,\ \ {\bf m}_f \ne {\bf m}_i \\ \nonumber &=& 
\sum\limits_{c,v} \  \frac{\mu_{ \bf{q}_c, \bf{q}_v}^\rho \ \mu_{\bf{q}_c, \bf{q}_v}^\rho}{(\hbar\omega_I + E_{i,{\bf m}} -E_{c,v,{\bf m}}  + i\Gamma )}, \ \ {\bf m}_f = {\bf m}_i \ \ \\ &&\textrm {(i.e. one gets Rayleigh scattering only)}
\end{eqnarray}
This is still 2$^{nd}$ order, but barren of Raman scattering. Something has to be done to create phonons.  Here we allowed the transition moments to vary, as they were always meant to, and are often allowed to do in other contexts, these past 90 years. The DR model left them frozen and went deeper into perturbation theory to generate phonons, which we believe are given a Pauli blocked still-birth with no path of the electron to return to the hole.  


         \vskip .07in
  \noindent  \subsection{Off-resonant  Raman scattering}
  \vskip .07in

  Off-resonance, the effective lifetime in the virtual excited conduction band states is $\hbar$ over twice photon energy gap $\Delta E$ of the laser promotion to  electronic resonance\cite{heller2,soo}. For very short times well off-resonance,   the  transition moment is applied twice to the initial state as the electron is promoted and then fills the hole; nothing more happens. There is not  time for nuclear wave packet motion  in  cases where equilibrium geometry changes in the excited state.  In the time domain picture of KHD,  wave packet motion takes  place along Born-Oppenheimer potentials mostly in steepest descent directions\cite{heller2}, electrons are interacting with phonons but most decidedly not inelastically.  The displaced phonon wave packets   are instantly registered as phonons as the excited state is reached, but the corresponding Raman intensity develops slowly -  first order in time  -  as the wave packet develops velocity\cite{heller2}.  In contrast, the transition moment coordinate dependence   means instant excited state phonon population   and Raman intensity.  This  ``$t^0$'' time dependance  is  more robust to off-resonant detuning, and starts to dominate far enough off-resonance\cite{heller2}.  Thus the  off-resonant D mode contribution suffers  even with impurities present  since there is not sufficient time  to backscatter, but G and 2D do well off-resonance since no backscattering is required.

\vskip .07 in
\subsection{Bandwidth outliers in figure~\ref{fig:width}}
  \vskip .07 in
  
  We now discuss the outliers marked by an asterisk seen  in figure~\ref{fig:width}.  The biggest deviations from the estimates are the D + G$^{\prime}$ band, just above 2D, about 50 cm$^{-1}$  broad as opposed to an estimate done with our assumptions of about 25 cm$^{-1}$, and the 2D + G$^\prime$, rarely reported experimentally at 4280 cm$^{-1}$ and about 80 cm$^{-1}$ broad\cite{rao2011multiphonon}. Unlike the other combination bands, the D + G$^{\prime}$ band requires impurities and elastic backscattering to be seen, allowing it to grow strong (and broad). 2D + G$^{\prime}$ does not require impurities. Presumably, different mechanisms are at work in each case, explaining why the bands do not fit the assumptions going into the bandwidth estimate and why D + G$^{\prime}$ needs backscattering.  

\begin{figure}[h] 
   \centering
   \includegraphics[width=5.0 in]{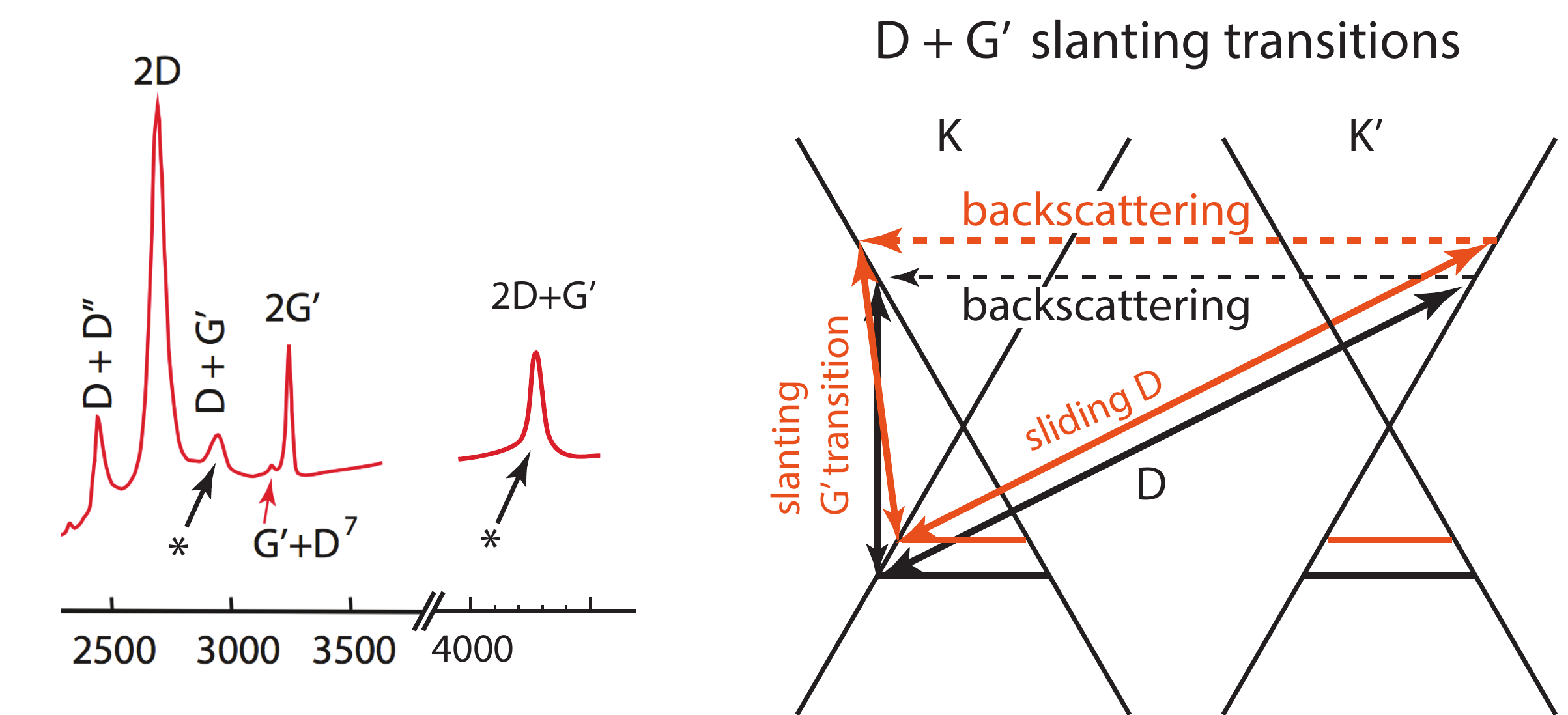} 
   \caption{Genesis of the 2950 cm$^{-1}$ band labeled, variously D+G (sometimes) or D+G$^\prime$(usually). It requires impurities to become visible. If D is produced by a sliding intervalley transition,  then after elastic backscattering the    G$^\prime$ transition   slants to reach the hole, generating a (generally) low $\bf q$    G$^\prime$ phonon  to preserve momentum.  The slanting transitions have somewhat   larger Raman shift than pure G;  the totality of  such transitions explains the   width of the 2950 cm$^{-1}$ band. The required backscattering  explains why impurities are required to make the D+G$^\prime$ combination overtone appear, unlike the other combination overtones, which can be faintly  seen in a relatively pure sample at the top right of figure 6 of the main text, $N_e =0$,  but are suppressed as impurities are added. (The  2D+ G$^\prime$  does not require impurities to be robust\cite{rao2011multiphonon}.) }
   \label{fig:D+G}
\end{figure}

Intervalley backscattering is necessary for an electron that produces a D (G$^{\prime}$) phonon upon excitement before it can emit a G$^{\prime}$ (D) phonon on the way down (see figure ~\ref{fig:D+G}). This is because the G phonon has a third of the unit cell of the D phonon, or equivalently, the D phonon is at the K point with respect to the G phonon. We further speculate that the nature of the scattering (figure ~\ref{fig:D+G})       allows the D+G band to be composed of both  D+G and G$^{\prime}$ in the following way: G does not participate in sliding, but it can ``slant'' (the analog of sliding, but for transitions near the vertical).   The G becomes a G$^{\prime}$ transition by becoming nonvertical (keeping the electronic energy fixed); this requires the creation of low $k$ G$^{\prime}$  phonons, to keep momentum conserved;  a   small $k^{\prime}$ deviation from the electronic $K+k $ is created, which with elastic backscattering of the electron becomes a   small $-k^{\prime}$ deviation from the $K-k $.  The next step, involving emission creating a  D phonon, varies in energy   according to   $-k^{\prime}$ and the D mode dispersion.

If G $^{\prime}$ is produced first, and slides, it is the D emission that must  slant on emission, causing a range of $k$ values for this dispersive band (although $k=0$ is forbidden by symmetry). 
Between these two possibilities,  there is easily a sufficient energy range of  phonons thus produced by slanting, accounting for both the enhancement of the ``D+G'' band intensity with backscattering defects (due to slanting)  and the breadth of the band. 

The same slanting transitions  can occur  when  producing lone G phonons in absorption, so why is the G band not correspondingly broadened? The reason is that the resulting low $k$ G$^{\prime}$  phonons are Pauli blocked, and elastic backscattering only blocks them further. The G bandwidth may however reflect the intrinsic ``Pauli blocking tolerance'' for very small $k^{\prime}$. 

The only three phonon band we discuss in this paper is G + 2D (or G$^\prime$ + 2D), which is   rarely reported experimentally at ca. 4270 cm$^{-1}$ Raman shift\cite{rao2011multiphonon}.  It is broad, with  roughly an 80 cm$^{-1}$ line width, and does not require impurity backscattering.  The reason for its existence and its linewidth have a plausible scenario  from our KHD based approach, including transition sliding. 
 
 The band may be produced by first creating a G$^\prime$ phonon in parallel with  a D in absorption, with a 2q momentum kick to the electron, followed by a -2q emission along the reverse D path creating a second D phonon.  The new twist here is that the first 2q kick can be shared in any proportion as 2q = 2q$_{\textrm{G}^\prime}$ + 2q$_{\textrm{D}}$.   Again, even without a detailed calculation of intensity distribution and fall-off, it is clear that there is more than 80 cm$^{-1}$ energy difference available depending on the  ratio of G$^\prime$ to D in the first step.  The emission is through a normal D at momentum 2q. The 2q = 2q$_{\textrm{G}^\prime}$ + 2q$_{\textrm{D}}$ process can also happen in emission.

\vskip .07in
\subsection{The ``molecular approach''}  
\vskip .07in

  An  important prior and non-DR   perspective on Raman scattering in graphene has been   termed the ``molecular approach" \cite{TommasiniMol,molApproach}.  A molecular polarizability context  of the type familiar from off-resonance Raman scattering was used. Electron-phonon ``scattering''   plays no role. 
   The off-resonance Placzek polarizability derivatives for computing Raman intensities, like the experimental   Raman bands,  are  often quite similar to on-resonance spectra for the conjugated hydrocarbons, except for overall intensity. A finite Placzek polarizability derivative  requires  non-constant transition moments.  However, an approach restricted to off-resonance polarizability cannot be regarded as a complete theory for resonance Raman scattering in graphene, but it is a step in the right direction away from DR.
   
   Consistent  with its off-resonant character,  the polarizability approach is a near instantaneous picture, leaving little to no time for electron-phonon scattering in the excited state.   The D mode  requires elastic backscattering (and some time) to become visible in the Raman spectrum, and indeed the polarizability picture is most successful it seems with the G and 2D modes, which require no backscattering.  
 The molecular polarizability picture is  a big step in the right direction,   quite distinct from  DR methods.   

Another interesting  and  instructive   contribution  to the molecular approach is found in Tommasini  \textit{et.al.} \cite{tommasiniMolecular}, attempting a more general electronic resonance formulation.  It was based on a KHD foundation but still resorted to the Condon approximation, i.e. constant transition moments.

\end{document}